\documentclass[12pt]{article}
\usepackage[utf8]{inputenc}
\usepackage{epigraph}
\usepackage{bbm}
\usepackage{lettrine}
\usepackage{tikz-cd}
\usepackage{amsthm,amsmath,amssymb}
\usepackage{graphicx}
\usepackage{subfigure}
\usepackage{changes}
\usepackage{soul}
\usepackage[T1]{fontenc}
\usepackage[affil-it]{authblk}
\usepackage{tikz}
\usepackage{float}
\usepackage{tikz-cd}
\usetikzlibrary{arrows}
\usepackage{scalefnt}
\usepackage{makecell}
\usepackage[nomessages]{fp}
\usepackage{enumerate}
\usepackage{tikz-qtree}
\usetikzlibrary{shapes,positioning,chains,fit,calc}
\usepackage{rotating}
\usepackage{cancel}
\usepackage{algorithm}
\usepackage{algpseudocode}
\usepackage{comment}
\usepackage{multirow}
\usepackage{caption}
\usepackage{relsize}
\usepackage{pagecolor,lipsum}%
\usepackage{fullpage}
\usepackage[colorlinks = true, linkcolor = blue,urlcolor  = blue,           citecolor = cyan,
anchorcolor = blue]{hyperref}
\usepackage[numbers,sort&compress]{natbib}
\usepackage{algorithm}
\usepackage[section]{placeins}

\usepackage{xspace}

\definechangesauthor[name={Ilya}, color=red]{is}
\definechangesauthor[name={Varsha}, color=blue]{vc}

\usetikzlibrary{shapes,positioning,chains,fit,calc}
\usetikzlibrary{decorations.markings}
\tikzstyle arrowstyle=[scale=1]
\tikzstyle directed=[postaction={decorate,decoration={markings,
		mark=at position .65 with {\arrow[arrowstyle]{stealth}}}}]
\tikzstyle reverse directed=[postaction={decorate,decoration={markings,
		mark=at position .65 with {\arrowreversed[arrowstyle]{stealth};}}}]

\usepackage{caption} 

\begin{document}

\title{Multiscale Planar Graph Generation    }
   \author[1]{Varsha Chauhan}
\author[2]{Alexander Gutfraind}
\author[1]{Ilya Safro}
\affil[1]{School of Computing, Clemson University, Clemson SC, USA\\
  \texttt {\{varshac,isafro\}@clemson.edu}}
\affil[2]{Loyola University Medical Center, Maywood IL, USA\\  \texttt{agutfraind.research@gmail.com}}
\maketitle
\section{Introduction}
A network is a representation of a set of entities and the relationships between them. The network paradigm is often used to represent physical, biological, engineered and social systems \cite{newman2018networks}. Networks can help us better understand the structural and functional dynamics of these systems and formulate predictive models. 
However, collecting real-world network data often requires time and can be expensive. Also, for many applications, the sensitivity of real-world data towards theft and misuse further adds to the cost of protection and security of the data, which sharply limits its availability. 

The problem of data scarcity can be tackled by using synthetic data which can mimic both the properties and diversity of real world networks.  Such synthetic data can be used for simulations, analysis, and performance/quality verification of algorithms - a crucial task in algorithm engineering. Synthetic network generation is one of the most important fields in network science from both theoretical and practical perspectives. We refer the reader for an in-depth discussion to recent reviews in \cite{gutfraind2015multiscale,staudt17generating-ans}.  

\subsection{Planar Graph Generation}
Planar graphs are the class of graphs that can be embedded in a two-dimensional plane without edge crossings. 
Designing efficient algorithms for planar graphs is an important subfield in the area of algorithm development and optimization \cite{meinert2011experimental}. From the practical perspective, the planarity is also an important characteristic of many  physical networks such as roads, utilities, water distribution systems, and some circuit designs. Many of these networks are, in fact, almost planar, that is, one can remove typically small fraction of edges to make them exactly planar. 

The wide range of real-world applications of planar networked systems has created a demand for planar graph generators. Although the planar graphs share the property that they can be embedded in a plane, a planar graph generator should also be able to replicate other properties exhibited in a real-world networks. \emph{However, the currently available synthetic network generators can either generate networks that exhibit realism with no planarity guarantees, or give planar networks with otherwise random structure that lack the structural characteristics of real-world networks.} Also, most of the existing research in general purpose network generation covers models related to scale-free networks, heavy-tailed degree distributions, and relatively high clustering coefficient that are not typical to real-world (almost) planar networks.

\subsection{Our Contribution}
In this paper, we present a flexible algorithm that can synthesize realistic planar replicas of a known planar graph that can be rescaled to much larger graphs. The method 
follows the multi-scale editing 
approach \cite{gutfraind2015multiscale} in which a given graph is projected into a hierarchy of its coarsened representations (coarse graphs) that are then perturbed by edits at various scales of coarseness in the hierarchy. The method preserves the structural properties including the planarity with \emph{controllable} bias, while introducing realistic variability at multiple scales of coarseness. Because the method belongs to the family of multiscale editing approaches, it generates planar graphs that attempt to replicate properties of the original graph at all levels of its coarse-grained resolutions which is the main property of the multiscale editing approach.

Throughout this paper we refer to the term ``realistic'' network multiple times. Realism of a generated similar network is not a uniquely defined notion as its meaning obviously depends on the application in which generating a similar to the original network is required. The question of realism definition is beyond the scope of this paper. We refer the reader to a discussion in \cite{gutfraind2015multiscale} and its preliminary extended version  \cite{gutfraind2012multiscale}. Intuitively, the multiscale generative method suggests that a realistic network is the one that replicates some properties of the original network at multiple scales of coarseness (in contrast to many different methods that generate similar networks with predefined properties such as clustering coefficient and degree distribution only at the finest scale). We advocate that preserving them at multiple scales is at least as important for a variety of applications as at the finest scale. Technically, in many cases, preserving just a couple of such properties as the degree and second shortest distance distributions, will imply preservation of many path-based metrics such as betweenness and diameter which does not necessarily happen at the finest level only methods. 


\section{Network Generation Algorithms} 

The field of network science and, in particular, network synthesis is vast and cannot be comprehensively reviewed here. Hence, we focus on several particularly illuminating approaches for modeling realistic networks that presumably may be applied as or changed to the first step in realistic planar graph generator. In contrast to the different versions of random planar graph generators, there is an obvious lack \cite{barthelemy2011spatial} of planar graph generators that generate graphs that are similar to the original planar graphs. This is the reason, why practitioners and decision makers use other graph generators in combination with planarization postprocessing to generate planar and hopefully realistic graphs. This is also a reason for our comparison with these algorithms in the next sections. These approaches fall into categories, namely, sampling models, generative models and editing models.
\subsection{Sampling models}
Sampling models are typically used for large scale networks. In this technique we pick a subset of vertices and/or edges from original graph and calculate the distribution of various graph properties such as degree distribution or link probabilities. The network is then generated by sampling from estimated distribution. One of the important examples of this model is the {\bf Exponential Random Graph Models (ERGM)} model.

The ERGM models \cite{hunter2008ergm} are a class of statistical models, earlier called p-star models, that are popular in the study of large-scale social networks. To build a network, the ERGM first estimates certain parameters by fitting an observed social network and then constructs new networks by sampling from the estimated distribution. For example, in the Bernoulli and Erdös-Rényi ERGM models which generate homogeneous networks, the parameter space is based on same probabilities for each added connection, whereas the Chung-Lu ERGM model \cite{aiello2000random} for large random graph with given degree distribution, the probability of connection of two nodes is proportional to the product of the degree of the nodes. The model can generate large graphs which depict some of the behaviors of massive realistic graphs and also predict the size and number of large components in the graph.
ERGM models are successful in generating social networks and exhibit realistic degree distributions and small world structures. Also there are several ERGMs with community structure \cite{karrer2011stochastic}, \cite{fronczak2013exponential},\cite{van2019reconstructing} but none of them give any planarity guarantees, and normally violate planarity. While potentially, this model could serve as the first step in planar network generation (the planarity could be one of the properties or it can be applied with subsequent planarization of synthesized network), we emphasize that it is extremely slow and cannot be applied even on medium size networks, so we cannot experiment with it and compare to our generator.\\

\subsection{Generative models} Generative models typically construct a network starting with an empty or small seed network and then iteratively add network elements (such as nodes and edges) to match some properties of a network that have to be preserved. These algorithms attempt to preserve the real network properties over the evolution and growth of the synthetic network. Important examples of generative models are the following.\\

\noindent {\bf BTER} Block Two-Level Erdös-Rényi model (BTER) \cite{seshadhri2012community} is based on the idea that a network contains communities that are Erdös-Rényi graphs in which each pair of vertices is independently connected with some probability. BTER graphs contain dense Erdös-Rényi communities that are found in real-world networks. The algorithm is two-phased. In the first phase, a collection of blocks or Erdös-Rényi  communities with specified degree distribution is created. Then the blocks made interconnected and excess degree nodes are removed based on Chung-Lu (CL) model \cite {aiello2001random} such that each subnetwork  is well modeled by CL. 
BTER has been shown to model realistically a variety of network properties, but as with ERGM, it gives no guarantees of planarity. Also, whether communities in (almost) planar networks have hierarchical and connectedness structure similar to BTER model or not is not explored.\\ 

\noindent {\bf RMAT and Stochastic Kronecker Graphs}
The Recursive Matrix graph generator introduced by Chakrabarti et al. \cite{chakrabarti2004r} and its extensions AutoMAT-fast \cite{chakrabarti2004r} can generate large-scale complex realistic networks. The generator is based on a recursive algorithm that operates on the adjacency matrix of the graph by dividing it into four equal-sized partitions and distributing edges to each partition based on fitting a set of parameters.\\ 
The  Stochastic Kronecker Graphs (SKG) \cite{mahdian2007stochastic} extends the methods of RMAT.  Similarly to RMAT it is a recursive model, which starts with a small initiator matrix and recursively produces large graphs by applying Kronecker products. SKG can be interpreted as network which is a hierarchy of communities which grow recursively to create copies of themselves and every node has unique set of attributes values. The model can generate graphs with static patterns such as degree distribution as well as temporal patterns such as diameter over time. As before, planarity is not guaranteed as well as the community structure similarity with real-world networks that have one.\\
    
\noindent {\bf Multifractal Network Generator}
In 2010, Pallaa et al. \cite{palla2010multifractal} introduced the multifractal network generator  which can generate realistic networks with specified statistical features. The method starts with defining a generative measure on a single fractal or unit square and calculating link probability. The network is then  scaled to the infinite limit by recursively dividing the fractal into a number of rectangles and introducing connections between them based on the link probability.  Although this method was able to generate small scale realistic graphs the recursive method was slow for large complex networks.  It is unknown if the generated networks can be constructed to have planar or quasi-planar structure, but the random nature of the construction suggests that planarity would be uncommon even in small graphs. However, the backbone networks generated by this model could be planar and thus possibly relevant to some infrastructure networks (for example, see major gas pipes in \cite{Newman2010}). Unfortunately, these networks have layers of fractals and do not exhibit properties of infrastructure networks such as small diameters, shortcut edges and redundancy in paths. Thus making comparison of these networks with infrastructure networks impossible.

\subsection{Editing models} 
The editing models approach starts with a given (real or empirical) network and controllably introduces random changes to its elements (such as nodes and edges) until the network becomes sufficiently different from the original network. These changes are designed to introduce variability while preserving key structural properties during the random editing.  Such methods are a promising direction for a relatively more realistic modeling of networks, and that includes properties such as planarity or near-planarity.\\

\noindent {\bf Edge-swapping}
The edge-swapping method \cite{Tabourier:2011:GCR:1963190.2063515,rao1996markov} is perhaps the first important algorithm in the class of editing models, and it is based on the insight that the degree distribution of a graph is preserved under a chain of edge-swapping operations.  Such a chain of edge swaps can even asymptotically achieve important mixing properties giving high variability.  Despite these successes, edge-swapping operations can be very disruptive to planarity and other global properties of the graph, and there are no good post-selection methods for achieving planarity.\\

\noindent {\bf Multiscale Network Generation}
In \cite{gutfraind2015multiscale}, several of us proposed a strategy termed MUSKETEER (Multiscale Entropic Network Generator) for realistic graph generation. The main idea was  based on the observation that the properties of real networks that should be preserved during generation are not only those measured at the finest resolution but also those that can be measured at the coarse resolutions. Multiscale generation leverages coarsening schemes used in highly-accurate multiscale solvers for combinatorial optimization such as linear arrangement, compression and partitioning \cite{safro:relaxml,safro:mlvsp,SafroRB08,safro2006graph,SafroT11}. In such coarsening schemes, nodes in a network are assigned into aggregates (or, typically, very small communities) which are themselves parts of  larger aggregates and so on in a hierarchical manner. The algorithm was successful in generating a number of replicas for several real-world original networks, but did not guarantee planarity. This paper continues this line of research and offers an implementation of the multiscale strategy that actually produces planar networks.\\

\noindent {\bf ReCoN} Staudt et al. \cite{staudt17generating-ans} later used  principles similar to those of multiscale method and developed a fast network generator that could generate large-scale replicas of real complex network that are structurally similar to original network. Instead of leveraging multiscale coarsening schemes, ReCoN generated synthetic networks by randomizing the edges between communities which were detected by the community detection methods while keeping the same degrees of nodes. ReCoN is built on top of the LFR generator implemented in \cite{staudt2014networkit}. 

\subsection{Planar Network Generators}
Planar networks with underling graphs have attracted a lot of attention since a landmark paper by Tutte \cite{tutte1963census}. Most of the research was dedicated on the study of structural properties (including their generation) of random planar graphs or uniform random planar graphs such as triangulations, and meshes. However, the currently available planar graph generators usually generate uniform random graphs by interpolation of planar subgraphs or generate planar subgraphs of a non-planar graph. Unfortunately, they are very far from being practically important for such tasks as generating graphs underlying infrastructure networks since they fail to present most other properties that  are viewed as significant in this area, such as the degree distribution, the community structure and others. Some important available planar graph generators are discussed below.\\

\noindent {\bf Plantri and Fullgen software}. Plantri \cite{brinkmann2007fast} can generate triangulations,  quadrangulations,  and  convex  polytopes using recursive algorithm which is efficient and fast. Fullgen \cite{brinkmann2011program} generates fullereness which are planar cubic graphs with $5$ or $6$ faces. The important characteristic of this software is that it generates only one graph as output from a family of isomorphic graphs saving the space needed to store them. The software also offers the user the option to restrict adjacent pentagons using an input parameter.  \\

\noindent {\bf Markov Chain Planar Graph Generator}. This algorithm was proposed by Denise et al. \cite{denise1996random} and is based on Markov Chain that generates planar subgraphs from a non-planar graph. The algorithm defines a Markov Chain on the state space of all subgraphs of the original graph and transitions as follows.  If an edge exists in space, it is deleted. If it is not present it is added in case it maintains planarity otherwise it is discarded. The method can successfully generate a planar subgraph in polynomial time. 
\\

\noindent {\bf Delaunay Triangulation and refinement method}. This method has been widely used by researchers to generate mesh networks. In \cite{shewchuk1996triangle}, Shewchuk presented an implementation of $2$-dimensional constraint Delaunay triangulation and Ruppert's \cite{ruppert1995delaunay} Delaunay refinement algorithm for mesh generation. \\

\noindent{\bf Geometric graphs}. Gilbert \cite{gilbert1961random} proposed a model to construct  random plane networks by first selecting points in infinite plane based on Poisson process with a specific density and then connecting points based on their distance (a parameter) from each other. The random geometric graphs closely represent the graphs generated by percolation process through various porous materials and therefore these graphs are extensively utilized by physicists to study continuum percolation models. Random geometric graphs also have application  in communication networks \cite{barthelemy2011spatial}.\\

\noindent{\bf Planar Erdös-Rényi graph}. In  1959,  Erdös and Rényi \cite{erdds1959random} introduced a method to generate a random graph with $N$ nodes and $m$ edges by connecting the edges randomly with independent probability $p$. The Erdos-Renyi planar graph generator generates random planar graph with uniform probability \cite{denise1996random} by rejecting the non planar edges thereby preserving planarity  \cite{denise1996random,gerke2004number, mcdiarmid2005random, barthelemy2011spatial}. This is the most basic planar model which cannot be directly used for practical replication purposes.

\subsection{Domain Specific Planar Network Generation}

Important applications of planar networks are infrastructure networks such as  roads, water distribution systems and power grids. There is a shortage of data for these networks owing to various reasons such as time and cost involved for collecting the data. Also the available data for infrastructure networks such as water distribution systems, and other utilities cannot be published due to confidentiality issues. As a result, the study of these networks and their simulation is highly dependent on the creation of high-fidelity synthetic data. 

In \cite{cura2015streetgen}, Cura et al. proposed a unified framework called StreetGen that works on real Geographic Information System (GIS) data and modeling hypothesis which automatized street reconstruction and generated a street network model which was coherent to real-world street model. StreetGen required parameters for specific street objects and needed specialization for different objects.

In order to generate a simulation data for grid networks, Wang et al.  \cite{wang2008generating} proposed an algorithm that generates random but realistic topology power grid networks that could be used as test power grids. The generator used probability distribution for defining for number of nodal locations, then the parameters for distance was used to generate simple topology which was connected. 


There is a similar shortage for data on water distribution systems (WDS) and the researchers typically need to rely on synthetic data to run simulation, test hypothesis and decision-making. The earlier methods for synthetic network generation involved manually creating realistic networks based on real data. Sitzenfrei \cite{sitzenfrei2013automatic} developed a software package DynaVIBe-Web that automated the generation of synthetic WDS networks.  The generator used street networks, GIS data and real data from more than one network. 
Murano et al. \cite{muranho2012waternetgen} developed an interactive application WaterNetGen as an extension to well-known WDS optimization tool EPANET \cite{rossman2000epanet}, which could generate network topologies, which is based on user-defined parameters and constraints.

Though, the above network generators produce realistic networks, they are specialized to specific domain and do not follow a generic approach to planar network generation. Also, replicating structural properties of a given network is not among the features of such generators. To our knowledge no work exists on cross-domain graph generators that generate realistic planar graphs that attempt to create graphs that are controllably similar to the given instances. 

\section{Notation}
Throughout the paper, we will use the notation $G=(V,E)$ for a graph, where $V$ is a set of $n$ nodes, and $E$ is a set of $m$ edges. We consider simple, undirected graphs, where $ij\in E$ is an edge connecting nodes $i$ and $j$, and the weight of $ij$ is denoted as $w(ij)$ and node volume (total weight of aggregated nodes) is denoted by $v(i)$. Both weight of edge and node are non-negative.(Although, our generator is not expected to work with weighted graphs, the weights will be used at the coarse levels to reflect aggregated nodes and edges.)

The subscript (such as $G_i$) used with a variable denotes the number of level in multiscale hierarchy. We denote an edited network in the hierarchy at level $i$ using superscript and subscript (such as $G^d_i$). A finally generated network at the level $i$ is denoted by $G^g_i$.

\section{Multiscale Planar Graph Generation}
Multiscale network generation (MNG) introduced in \cite{gutfraind2015multiscale}  is an editing model that generates realistic networks. The proposed multiscale planar graph generator follows the main ideas of MNG and makes them applicable on planar graphs. For the completeness of the paper we also overview basic components of the original method.

\begin{figure}[t]
\includegraphics[width=1 \textwidth]{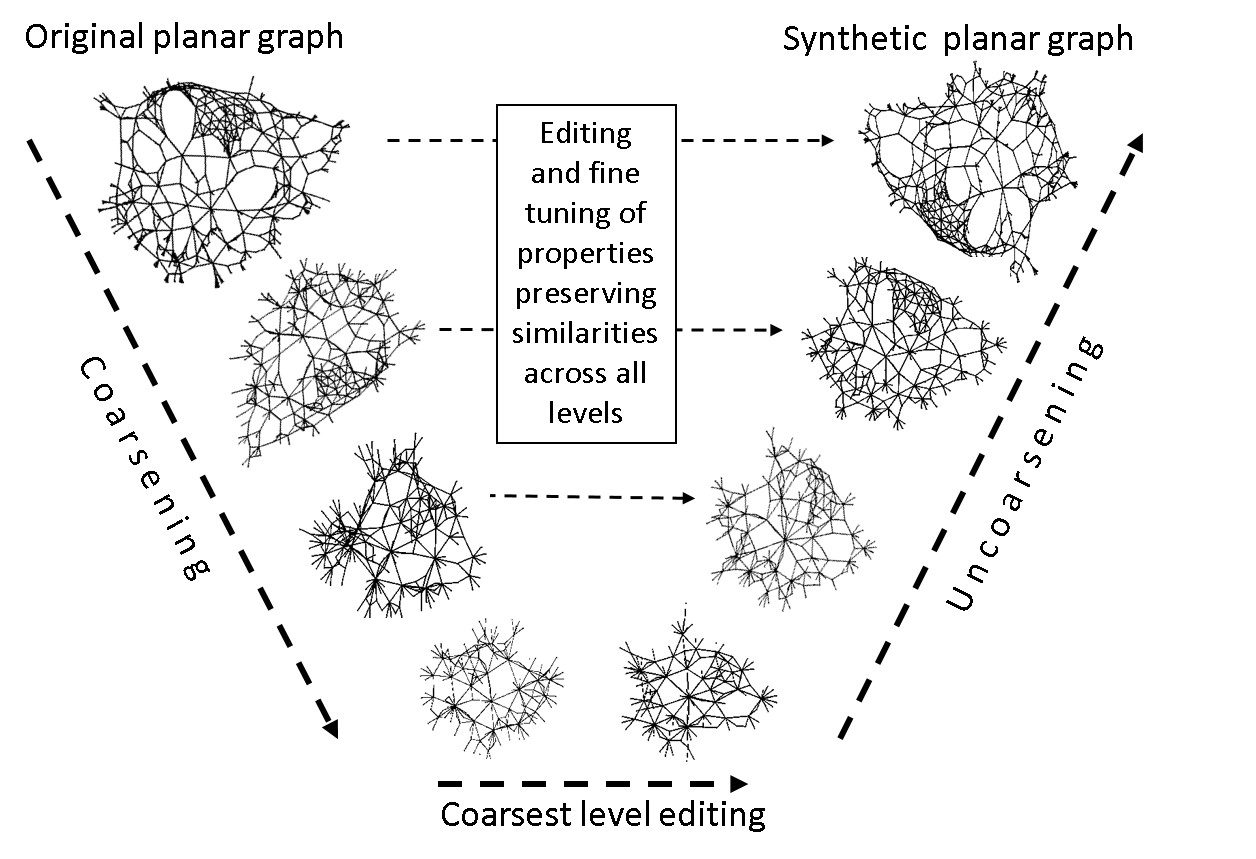}
\centering
\caption{The V-model for multiscale planar network generation. The original input planar network is coarsened to generate a hierarchy of coarse networks, the process is then reversed generating fine-level networks. The number of level (here 5) depends on the size of input network or the user input.}\label{fig:Figure 1}
\end{figure}

MNG follows a multilevel coarsening/uncoarsening scheme shown in Figure \ref{fig:Figure 1}. We start with an input graph $G$ and generate a hierarchy of next coarser graphs, $G_0=G, G_1, ... , G_k$, where $k$ is the number of coarsest level. The number of coarsened levels depends on the structure and size of $G$, and user input, which is a vector where each value determines the   required edit or growth rate at each level  of the hierarchy, and the length of vector determines the number of desired coarsening levels. The hierarchy construction (coarsening) is terminated if the graph is too small or too dense (density of graph $>0.9$) at some level (i.e., the coarsest level is reached). The  coarse  level construction is generic and based on the weighted aggregation method for combinatorial optimization problems \cite{safro2006graph,safro:relaxml}. Currently, it does not depend on the application predefined aggregates  in the network such as knowledge about real communities. However, this process can be adjusted as we did in \cite{staudt2016generating}. 

In order to generate a synthetic graph, during the uncoarsening stage, we introduce a series of local randomizations at different levels whose amounts can be specified by user's input. As mentioned previously, the user can control the number of levels coarsening and amount of edits or perturbations at each of these levels using parameters edge edit rate and edge growth rate. If user is interested in only local changes without destroying the global structure of the network, only fine levels are specified for randomizations. Otherwise, any realistic changes in global structure will require randomizations at coarse levels. During the uncoarsening, these randomizations are carried forward to the next finer level in the hierarchy. In Algorithm \ref{alg:MPNG}, we describe the sequence of steps in generating planar graph. We will now discuss each phase and notation in detail and our approach to generate planar graphs using the multiscale method.\

\begin{algorithm} 
\caption{Multiscale Planar Network Generator MPNG($G_i$)} \label{alg:MPNG}
\begin{algorithmic}[1]
\If {$G_i$ is not small or too dense or perturbations are required for $G_i$ at level $i$ by user}
    \State $G_{i+1} \gets$ create aggregated network from $G_i$ \Comment{see Alg. \ref{alg:coarsening}}
   \State $G_{i+1}^g \gets$ MPNG($G_{i+1}$) \Comment{Return coarser edited network from recursive call}
    \State $G_i^{d'} \gets $  interpolateUneditedAggregates from $G_{i+1}^g$
    \State $G_{i}^{d} \gets$ interpolateEditedAggregates($G_{i+1}^g$,$G_{i+1}^{d'}$)
\EndIf
\State $Q_{i} \gets $ measure properties of $G_i$
\State  $G_i^g \gets $ editing $G_i^d $ preserving $Q_i$
\State Return $G_i^g$
\end{algorithmic}
\end{algorithm}

\subsection{Coarsening} 
Since the input graph $G_0$ is planar, the aggregation algorithm that creates  coarsened graphs $G_i$ makes them also planar, so we follow the same coarsening scheme as that in the original MNG. Algorithm \ref{alg:coarsening} lists the steps involved for generating coarse level graph $G_{i+1}$ from $G_i$. 
\begin{algorithm} 
\caption{Coarsening($G_i$)}\label{alg:coarsening} 
\begin{algorithmic}[1]
\If { $G_i$ is not the coarsest graph }  
    \State Find set of seed nodes ($C$) for coarse network $G_{i+1}$
    \State Find fine-level nodes that belong to each aggregate
    \State Calculate weight of edges connecting aggregates and weights of coarse nodes 
\State Return $G_{i+1}$
\EndIf
\end{algorithmic}
\end{algorithm}\\
At each level of the hierarchy, we start with finding set of seed nodes, $C$, and its complement fine-level nodes $F$ which is based on two rules: (a) nodes with high volume and connectivity (i.e., major aggregates) are more likely to be included in $C$, and (b) the remaining nodes in $F$ should be ``strongly'' coupled to enough neighbors in $C$. At level $i$, given a graph $G_i$, to generate coarse level nodes for $G_{i+1}$ we begin with $C=\emptyset$ and $F = V_i$ where $V_i$ is set of nodes at fine level $G_i$. Next, we iteratively  transfer some nodes from $F$ to $C$ (visiting them one after another in random order), such that currently visited node $i\in F$ is added to $C$ if it is not well connected to those already chosen to $C$  \cite{safro2006graph}. The connection strength between nodes $i$ and $j$   is determined by means of normalized weight of edge $ij$ with respect to $C$, namely, if node $i\in F$ is not connected strongly enough to the currently chosen $C$, i.e.,
\begin{equation}\label{eq:cnodeadd}
\frac{\sum_{j\in C} w(ij)}{\sum_{j\in V} w(ij)} \leq \alpha,
\end{equation}
then we move $i$ to $C$ and transfer our attention to the next node in $F$. Thus, instead of requiring for a certain number of $F$-nodes to be transferred to $C$, we scan them iteratively, and decide based on Eq. (\ref{eq:cnodeadd}). The connection strength is parametrized using threshold $\alpha$ which determines the speed and number of coarse nodes (and implicitly edges). Big values of $\alpha$ (in multiscale algorithms, typically, 0.7 or bigger) will result in small changes in coarsened graphs that are created from level to level as most of nodes wont be strongly connected to $C$ (according to Eq. \ref{eq:cnodeadd}). In contrast, small values (in multiscale algorithms, typically, 0.3 or smaller) will cause a decreased number of levels as not too many nodes will be transferred to $C$. In all our experiments we have used $\alpha$ as $0.5$, which guarantees uniform coarsening. However, we note that the strength of connectivity criterion in network generation requires further investigation similar to that in multiscale optimization \cite{vlsicad} where it plays a crucial role in the solution quality.

The final phase of coarsening is computing the connection strength between the coarse nodes. Here we define the algebraic multigrid interpolation matrix  $P$ of size $ |V|\times|C|$ (for details see \cite{safro2006graph}) in which $P_{uv}$ represents the likelihood of $u\in V_i$ to belong to the aggregate seeded with node $v\in V_i$. The Laplacian of the coarse graph $G_{i+1}$, $L_{i+1}$, can be calculated by the algebraic multigrid coarsening operator $L_{i+1} \gets P^{T}L_{i}P$ where, $L_i$ is the Laplacian of $i$th level graph, and 
\begin{equation}
P_{uv} =
\begin{cases}
1, & \text{for }u \in C, v=u \\
0, & \text{otherwise}.
\end{cases}
\end{equation}

The edge $st$ connecting two coarse nodes $s\in V_{i+1}$ and $t\in V_{i+1}$, is assigned with the weight 
\[
\sum\nolimits_{k\ne l}P_{ks}w(kl)P_{lt}
\]
and the volume of the coarse aggregate seeded by $u\in V_i$ is $\sum\nolimits_jv(j)P_{ju}$.

This step finalizes creating the $(i+1)$th level graph, and we can measure the properties of $i$th level graph and store them in $Q_i$. As in the original MNG, we attempt to preserve the small loop structure of the network by avoiding any insertion of edges that connect nodes that were previously separated by the long distances in the original network. This is done by estimating the empirical probability of closed random walks (of limited length) that start at some node whose degree is at least two (for details see \cite{gutfraind2015multiscale}). In general, this step is application dependent as in different applications the preserved properties may vary. Because, in planar graphs of infrastructures it is important to generate realistic path lengths (e.g., not to create shortcuts that connect distant regions in a graph), we are sampling using random walks the distribution of path lengths and shortcuts (second shortest distance between nodes) and store them in $Q_i$. 

\subsection{Uncoarsening} 

Once the coarsest level is reached, we start the uncoarsening. During this process, at each level $i+1$ we choose nodes and edges to be edited (randomized while keeping some properties preserved), to generate edited network $G_{i+1}^g$ at level $i+1$ and then project the newly created graph to generate the next finer level $G_{i}^g$.

The projection is done in two steps. 
First, we interpolate the unedited aggregates (nodes and edges) in \textbf{interpolateUneditedAggregates} (see Alg. \ref{alg:interuned}) from $G_{i+1}^g$ to generate graph $G_i^{d'}$. This process is just a reverse interpolation of aggregates based on aggregation data stored during the coarsening phase. \emph{Because the input network at all levels is planar, the interpolated edges do not violate planarity.} This helps in preserving structural properties of original input network, as after this step we have a subgraph $G_i^{d'}$ of original network coarsened at level $i$. 

In the next step, we interpolate the edited aggregates in function \textbf{interpolateEditedAggregates} to generate graph $G_i^{d}$ by adding edited nodes and edges to graph generated with Algorithm \ref{alg:interuned}. The pseudocode for this step is presented in Algorithm \ref{alg:Interpolating Edited Aggregates}. We first interpolate the edited nodes and add edges that were trapped within the aggregates (or coarse) nodes connecting the interpolated fine nodes, i.e., these are edges that connect fine nodes that are coarsened within the same coarse node. Next we interpolate edited or new edges introduced during editing phase at level $i+1$. In this step, we introduce new fine level edges for every coarse edge connecting a pair of  aggregates $u$ and $v$. This is done by randomly selecting the fine level nodes generated by interpolating $u$ and $v$. The number of fine level edges added for each coarse edge is based on the degree distribution of nodes in aggregates $u$ and $v$ at level $i$ as stored in data structure during coarsening. This interpolation is likely to introduce crossing over edges, therefore, when we add an edge $ij$ to $G_{i}^d$, we check if the network is still planar.  If it is not, the edge is discarded. If an edge is discarded, we perform several iterations and find an edge which is similar to the edge $ij$ using properties stored in $Q_i$ during coarsening in Algorithm \ref{alg:MPNG} (i.e., in our implementation, the short loop structure).

\begin{algorithm} 
\caption{interpolateUneditedAggregates ($G_{i+1}^g$)} \label{alg:Interpolating UnEdited Aggregates}\label{alg:interuned} 
\begin{algorithmic}[1]
    \State $G_i^{d'} \gets $  uncoarsen nodes from $G_{i+1}^g$ using data stored  during coarsening at level $i$
     \State $G_i^{d'} \gets $ uncoarsen unedited edges $G_{i+1}^g$ using data stored during coarsening at level $i$
\State Return $G_i^{d'}$
\end{algorithmic}
\end{algorithm}

\begin{algorithm} 
\caption{interpolateEditedAggregates ($G_{i+1}^g$ , $G_{i}^{d'}$)} \label{alg:Interpolating Edited Aggregates} 
\begin{algorithmic}[1]
    \State $G_i^d \gets $  uncoarsen nodes from $G_{i+1}^g$
     \State $G_i^d \gets $ uncoarsen edited edges $G_{i+1}^g$
\State Return $G_i^{d}$
\end{algorithmic}
\end{algorithm}
After the interpolation is complete and we have a fine-level graph $G_i^d$, on which we introduce randomizations or editing (discussed below in detail) specified by the user at level $i$ to generate a finer-level random planar network network $G_{i}^g$. The topology of the final network depends on the level at which the changes are introduced and the number of edited network elements both dependent on user input. At the coarsest level, every network element is an aggregate which interpolate of many network elements at fine level, a small change introduced at this level may generate high-entropy changes which are carried forward to the next fine level, whereas addition of an element at fine levels may introduce elements to the final synthetic network. In general, the changes introduced to deeper levels of aggregation, the more significant changes are introduced in the topology. 

\subsection{Editing} 

In the final phase we measure the properties of the generated graph $G_i^d$ and compare with the properties of original graph $G_i$ coarsened at level $i$ which is stored in $Q_i$, thus preserving the local topological structure of the network and preventing addition of edges between nodes which were separated by long distance in original network at coarse level $G_i$ stored in $Q_i$. We then use an editing process which introduces randomizations in the network to generate a synthetic network. This is a process of deleting and adding new edges whose both amounts depend on the user input for level $i$ (namely, how much randomization is required at each level, i.e., a value from 0 - no randomization, to 1 - everything is randomized). 
In particular, we are interested in two properties, namely, the second shortest path length distribution (to prevent generating unrealistic shortcuts) and planarity. The first property, second shortest path length  $\textsf{spath}(u,v)$ for an edge $uv$ is the number of edges in the shortest path $u\rightarrow v$ that does not include edge $uv$. Before introducing edits we estimate the distribution of $\textsf{spath}$ at level $i$, $\mathbb{P}_{i}$ by random sampling of edges $R \subset E_i $ where $E_i$ is set of edges in graph $G_i$ as,
\begin{equation}
\mathbb{P}_{i}[d] \approx \frac{|{\{\{u,v}\in R : \textsf{spath}(u,v) = d\}|}{|{R}|}
\end{equation}
When we delete an edge $u_1v_1$ such that $\textsf{spath}(u_1,v_1) = d$, we choose a random node $u$ and $v$ at randomly drawn distance from $u$ using $\mathbb{P}_i$. This sampling of distances preserves multiple structural properties such as clustering coefficient and average shortest path (see more details in \cite{gutfraind2015multiscale}).

The second critical property is planarity. In order to preserve planarity, if inserting the new edge makes the network non-planar we discard it and find an alternate edge that preserves the desired structural properties (in this case the first property) as well as planarity. Technically, it is done by verification of existence of Kuratowski subgraph \cite{thomassen1981kuratowski} after adding a new edge. This step is repeated until we find a non-crossing edge that preserves the planarity of the network and thus generating synthetic planar graph $G_i^g$ at coarse level $i$.

\subsubsection{Rescaling} 

Rescaling is a part of the editing phase in which we add new elements (edges and nodes) to the synthetic network in order to increase its size. The scaling factor and the coarsened level at which the network is rescaled is controlled by node growth parameter which is provided as an input from the user depending on the user requirement. In general, rescaling at coarsest levels will preserve the local structure of the input network, i.e. the generated network will have increased number of communities whereas rescaling at finer levels will increase the size of communities. The scaling factor ranges from $0$  to any number which decides the percentage of new nodes that are to be added at the level $i$. This is a two step process. First, we introduce a new node $u$ and connect to an existing node $v$ in the network deleting an existing edge from $v$ to restore the degree of node $v$. In the next step, we find neighbors of $v$ iteratively over increasing distance from $v$ and connect the newly added node $u$ to the neighbors of node $v$ thus preserving the local topological structure of the network at coarse level $G_i$ stored in $Q_i$. The process is terminated when the desired number of network elements are added and a rescaled network $G_i^g$ is generated at coarse level $i$.

\section{Computational Experiments}
In this section we show the computational results summarizing the performance of our multiscale planar network generator in replicating the original and also generating rescaled networks. To test the variability of the generator we used real-world infrastructure networks such as water distribution system, power grid and road network that are either planar or have very few edge crossings that we removed. We used the water network from ``The Battle of the Water Networks II`` \cite{ostfeld2008battle} and for road network we used a sub graph of Texas \cite{leskovec2009community} road network from \cite{snapnets}. We also used a finite element large planar sub-graph of a finite-element graph from Boeing collection in \cite{Davis1997}. In case of the power grid \cite{snapnets} which was not completely planar, we generated approximate maximal planar subgraph of the network using Open Graph Drawing Framework (OGDF) \cite{chimani2013open} to be used as an input to our algorithm.

\subsection{Replication}
We tested our implementation on three sets of parameters, namely, ``Musketeer Coarse'' (at only two coarsest levels 5\% randomizations are allowed), ``Musketeer Fine'' (at only two finest levels 5\% randomizations are allowed), and ``Musketeer All'' (small 1\% randomizations are allowed at all levels). The amount of edits for each set of parameters are controlled such that the number of randomizations in final generated graphs for each set of parameters are comparable. Because randomizations and editing are introduced at all levels, even very little changes at the coarse levels will result in significant changes at the finest level in generated synthetic graph. 

We generated 30 network replicas for each network and compared the replicas with the original network based on the following metrics: number of nodes and edges, number of components, clustering coefficient, average degree, total degree-degree assortativity, average harmonic distance, modularity, pagerank and average betweenness centrality. We also compared our results with the existing generative models implemented in \cite{staudt2014networkit}, namely, ReCoN, RMAT and BTER and stochastic Kronecker graphs by generating replicas of same input network. Since, these models do not necessarily generate planar network, \emph{we post-processed the generated networks to find the maximal planar subgraph of the replicas} using OGDF library which uses edge removing technique, i.e., it adds one edge at a time while preserving planarity, if addition of the edge results in a non-planar graph then the edge is discarded thus generating a planar subgraph. We compared the generated planar graphs with the original graphs for the structural properties mentioned above. Clearly, one may argue that these generators were not developed to planar networks. We, however, note that these methods with planarization post-processing were chosen because there is no other acceptable solution to generate more or less realistically looking planar network that is similar to the original. As mentioned earlier, the available planar graph generators are generative models which either create specific classes of graphs with restricted values for minimum degree and connectivity (e.g., plantri and fullgen  or Delaunay triangulation methods) or generate random realistic spatial networks based on give probability $p$ (e.g., planar Erdos–Renyi, and  spatial Watts-Strogatz generator). Other examples include domain specific generators for road networks (e.g., StreetGen) and power grid random networks that are not necessarily planar networks. To the best of our knowledge, there is no domain independent generator whose goal is to preserve similarity with the input network.

\begin{figure}[t]
\includegraphics[width=1 \textwidth]{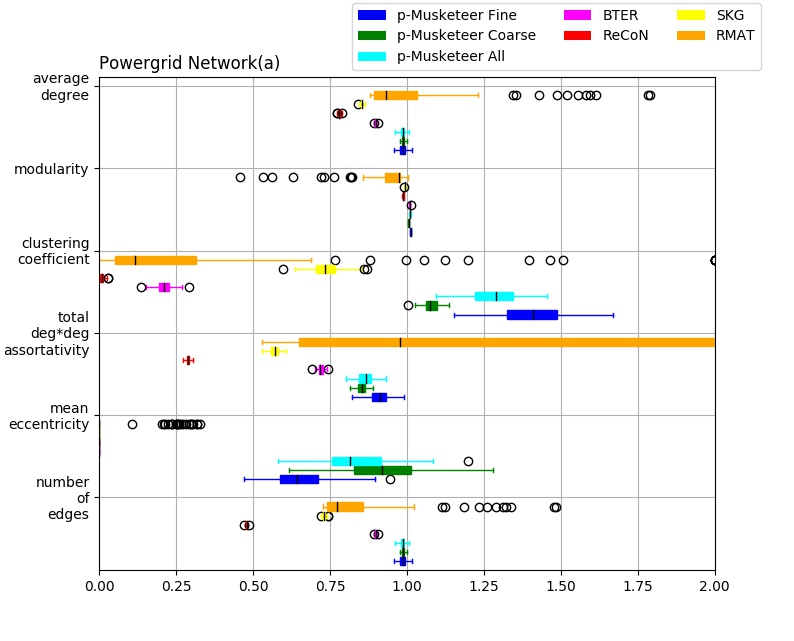}
\centering
\caption{Computational results on performance of planar Musketeer on power grid graph opsahl-powergrid with $4941$ nodes and $6211$ edges for clustering coefficient, number of edges, mean eccentricity, total degree *degree assortativity, modularity and average degree.}\label{fig:Figure 2}
\end{figure}

\begin{figure}[t]
\includegraphics[width=1 \textwidth]{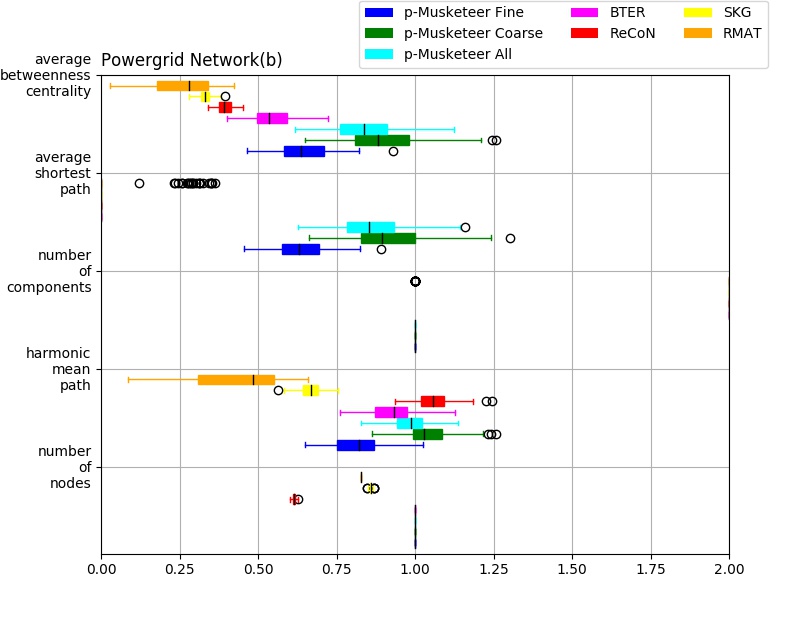}
\centering
\caption{Computational results on performance of planar Musketeer on power grid graph opsahl-powergrid with $4941$ nodes and $6211$ edges for number of nodes, harmonic mean path, number of components, average shortest path and average betweenness centrality.}\label{fig:Figure 3}
\end{figure}

\begin{figure}[t]\includegraphics[width=1 \textwidth]{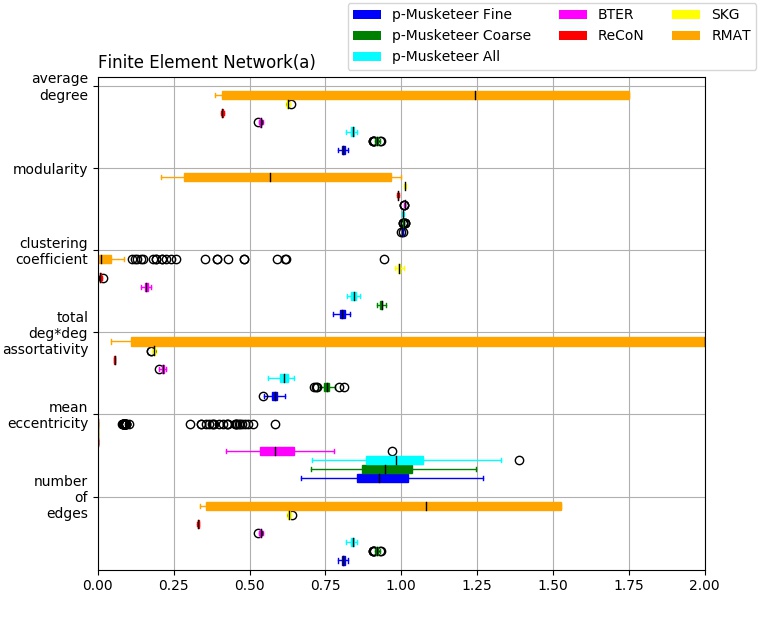}[t]
\centering

\caption{Computational results on performance of planar Musketeer on finite-element graph with $4704$ nodes and $13427$ edges for clustering coefficient, number of edges, mean eccentricity, total degree *degree assortativity, modularity and average degree.}\label{fig:Figure 4}
\end{figure}

\begin{figure}[t]
\includegraphics[width=1 \textwidth]{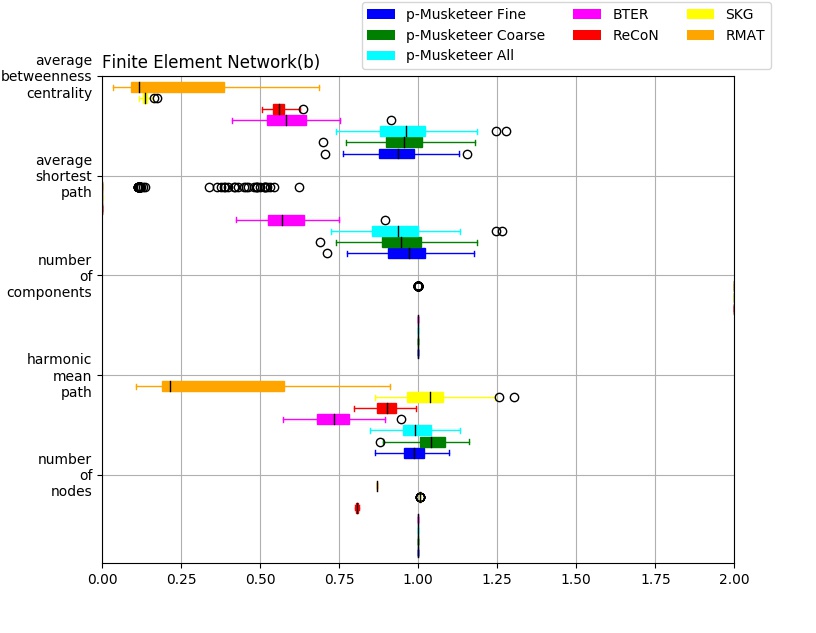}
\centering

\caption{Computational results on performance of planar Musketeer on finite-element graph with $4704$ nodes and $13427$ edges for number of nodes, harmonic mean path, number of components, average shortest path and average betweenness centrality.}\label{fig:Figure 5}
\end{figure}

\begin{figure}[t]
\includegraphics[width=1 \textwidth]{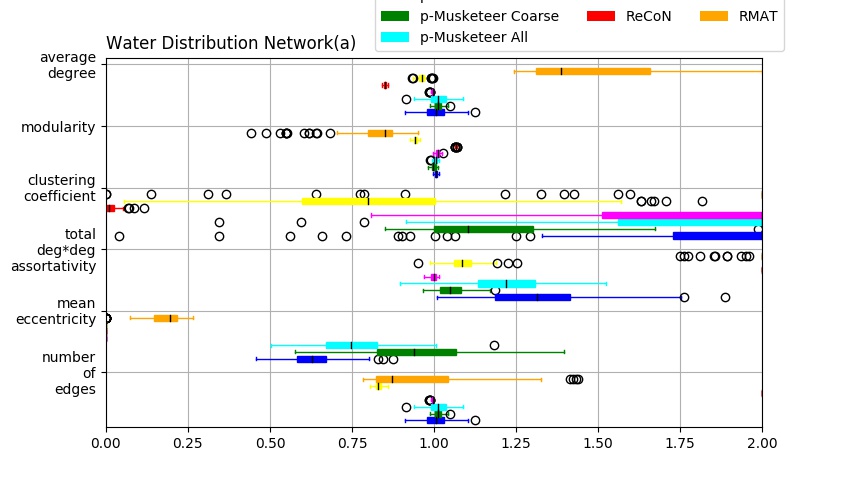}
\centering

\caption{Computational results on performance of planar Musketeer on real water network with $407$ nodes and $459$ edges for clustering coefficient, number of edges, mean eccentricity, total degree *degree assortativity, modularity and average degree.}\label{fig:Figure 6}
\end{figure}

\begin{figure}[t]
\includegraphics[width=1 \textwidth]{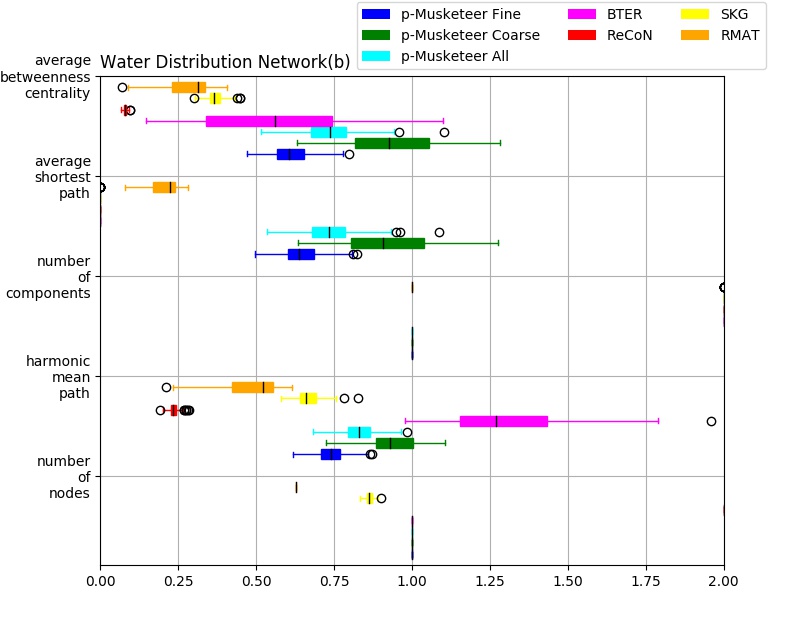}
\centering

\caption{Computational results on performance of planar Musketeer on real water network with $407$ nodes and $459$ edges for number of nodes, harmonic mean path, number of components, average shortest path and average betweenness centrality}\label{fig:Figure 7}
\end{figure}

\begin{figure}[t]
\includegraphics[width=1 \textwidth]{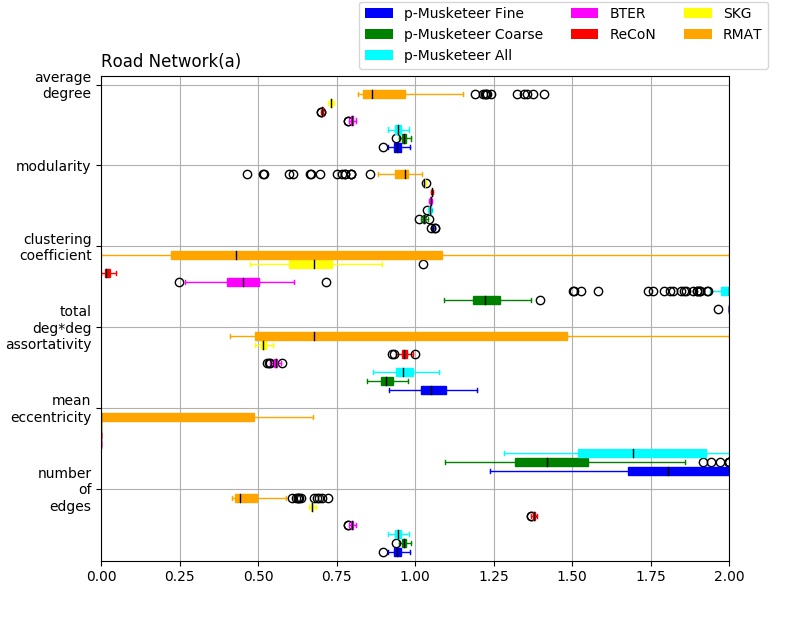}
\centering
\caption{Computational results on performance of planar Musketeer on road network from roadNet-TX with $2001$ nodes and $2957$ edges for clustering coefficient, number of edges, mean eccentricity, total degree *degree assortativity, modularity and average degree. }\label{fig:Figure 8}
\end{figure}

\begin{figure}[t]
\includegraphics[width=1 \textwidth]{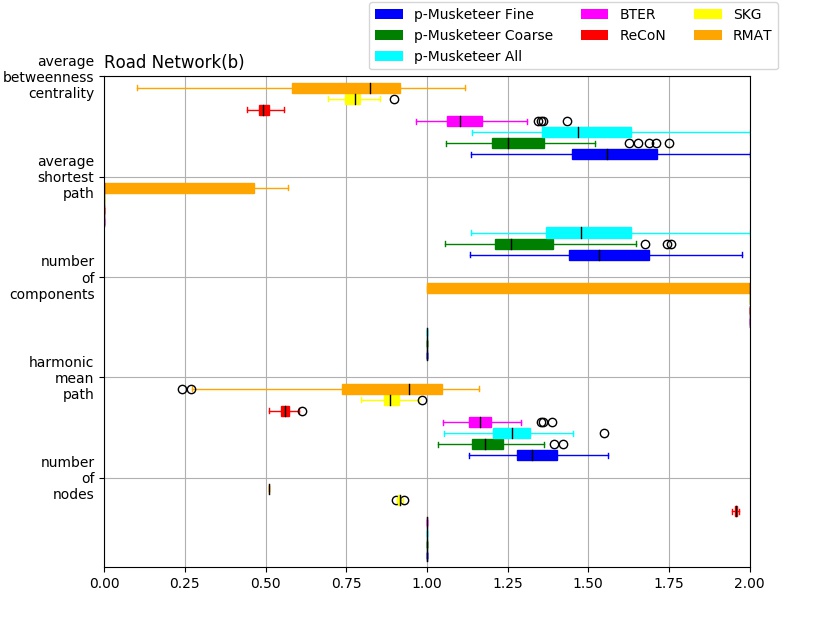}
\centering
\caption{Computational results on performance of planar Musketeer on road network from roadNet-TX with $2001$ nodes and $2957$ edges for number of nodes, harmonic mean path, number of components, average shortest path and average betweenness centrality. }\label{fig:Figure 9}
\end{figure}

The structural properties of the replicas were normalized such that $1$ denotes the property of original network. We performed $30$ experiments for each set of parameters the results for which is graphically represented in Figure \ref{fig:Figure 2}-\ref{fig:Figure 9}. 
 Our results indicate that multiscale planar graph generator can generate replicas that preserve almost all the properties of the original networks with relatively small deviation. Also, we observe that graphs generated by BTER and RMAT after planarization are close to original network (within $0-2$, where $1$ represents the property of original input network after normalization) for properties such as average degree and mean harmonic distance whereas the properties for networks generated by stochastic Kronecker graphs (SKG) are far from those in the original graphs. As such the plots for properties for the networks generated by SKG are not represented in the plots. However, we note that the distortion of properties on the replicas by other network generators may have been the result of the post-processing step (maximal planar sub-graph of the generated replica), which often created more than one connected components.
 


\begin{figure}[t]
\includegraphics[width=1 \textwidth]{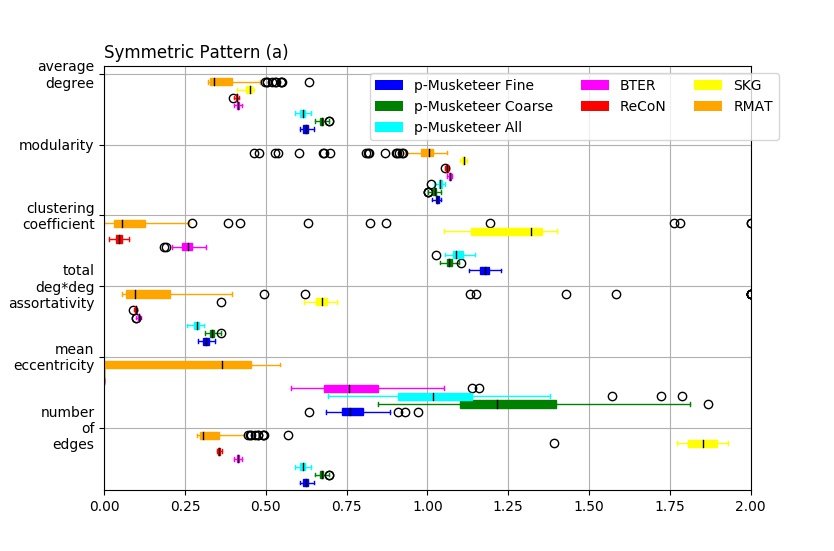}
\centering
\caption{Computational results on performance of planar Musketeer on symmetric pattern with $1141$ nodes and $3162$ edges for clustering coefficient, number of edges, mean eccentricity, total degree *degree assortativity, modularity and average degree. }\label{fig:Figure 10}
\end{figure}
\begin{figure}[t]
\includegraphics[width=1 \textwidth]{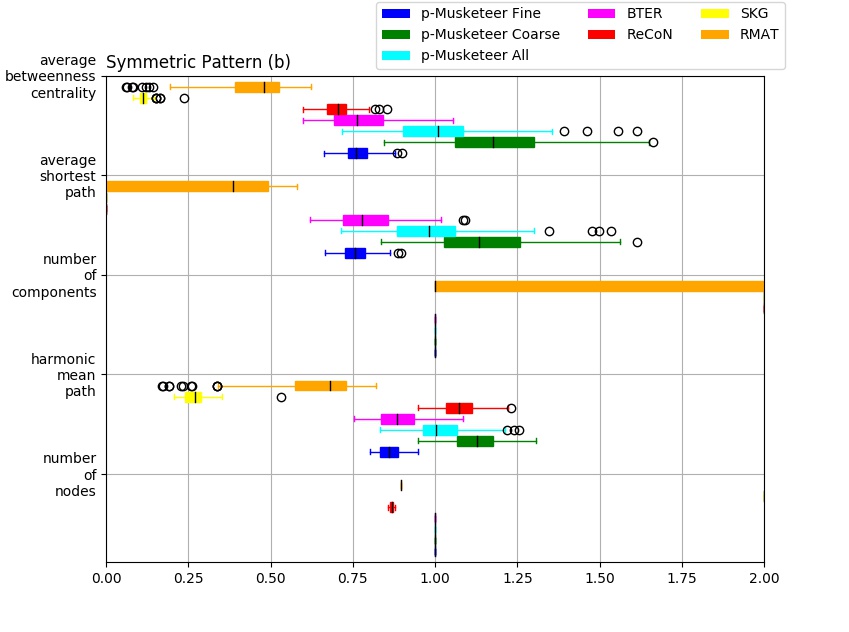}
\centering
\caption{Computational results on performance of planar Musketeer on symmetric pattern with $1141$ nodes and $3162$ edges for number of nodes, harmonic mean path, number of components, average shortest path and average betweenness centrality. }\label{fig:Figure 11}
\end{figure}
\begin{figure}[t]
\includegraphics[width=1 \textwidth]{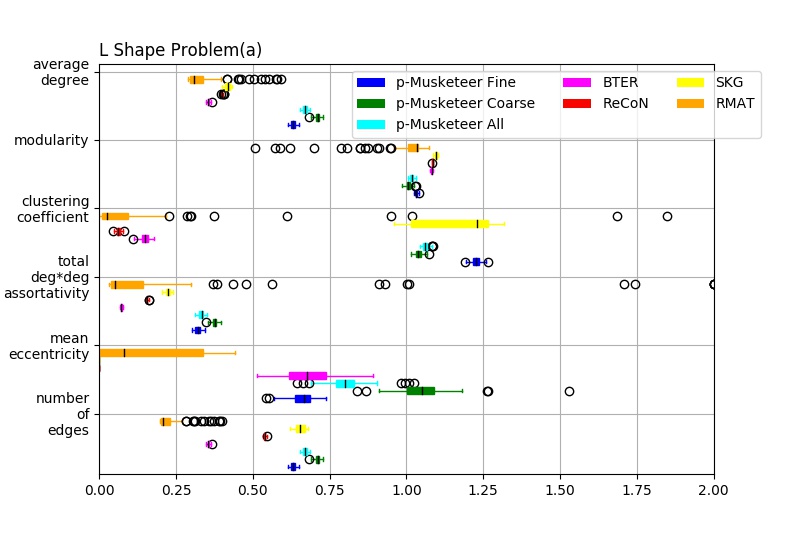}
\centering
\caption{Computational results on performance of planar Musketeer on graph for thermal L-Shape Problem with $3025$ nodes and $8904$ edges for clustering coefficient, number of edges, mean eccentricity, total degree *degree assortativity, modularity and average degree. }\label{fig:Figure 12}
\end{figure}
\begin{figure}[t]
\includegraphics[width=1 \textwidth]{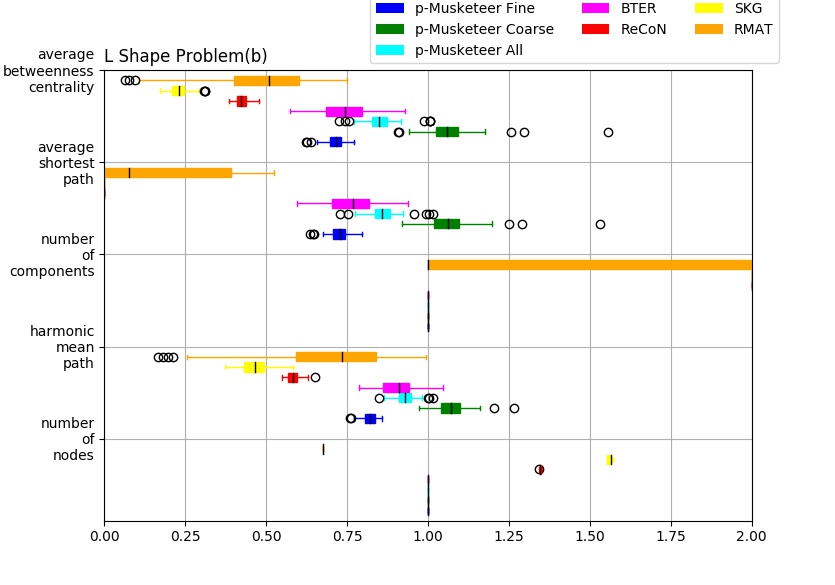}
\centering
\caption{Computational results on performance of planar Musketeer on graph for thermal L-Shape Problem with $3025$ nodes and $8904$ edges for number of nodes, harmonic mean path, number of components, average shortest path and average betweenness centrality. }\label{fig:Figure 13}
\end{figure}
\begin{figure}[t]
\includegraphics[width=1 \textwidth]{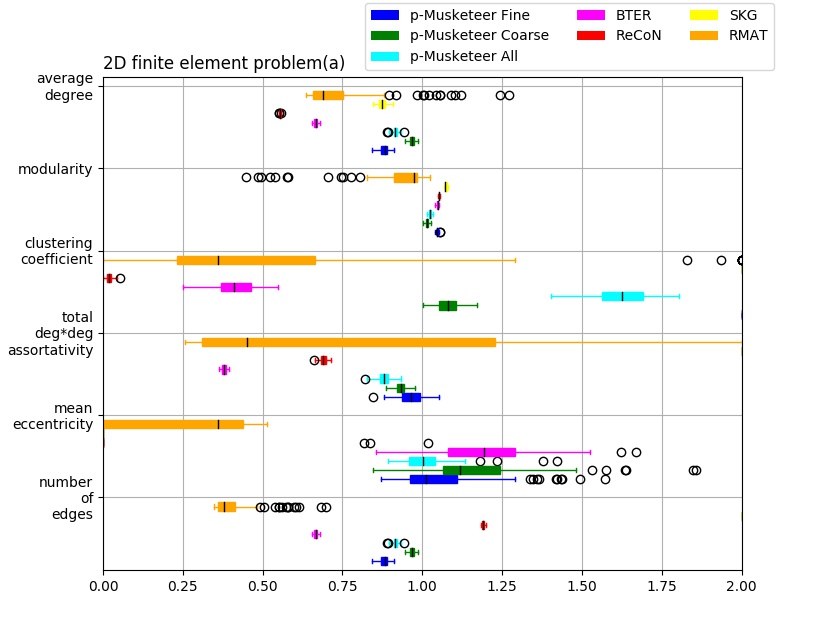}
\centering
\caption{Computational results on performance of planar Musketeer 2D finite element graph with $1866$ nodes and $3538$ edges for clustering coefficient, number of edges, mean eccentricity, total degree *degree assortativity, modularity and average degree. }\label{fig:Figure 14}
\end{figure}
\begin{figure}[t]
\includegraphics[width=1 \textwidth]{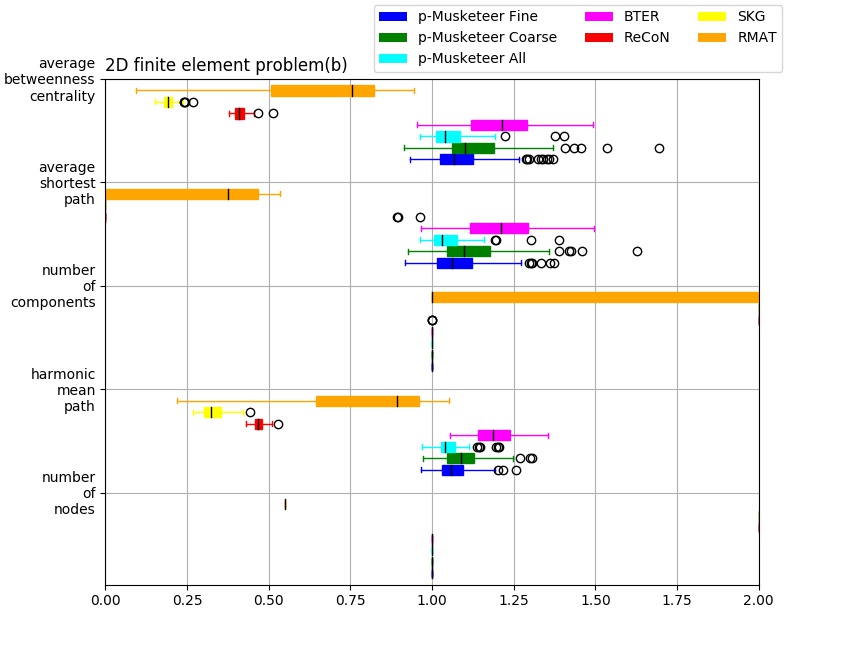}
\centering
\caption{Computational results on performance of planar Musketeer 2D finite element graph with $1866$ nodes and $3538$ edges for number of nodes, harmonic mean path, number of components, average shortest path and average betweenness centrality. }\label{fig:Figure 15}
\end{figure}
\subsection{Rescaling}

Our second set of experiments was designed to generate rescaled networks. We tested our implementation on three sets of parameters, namely, ``Musketeer Coarse'' (30\% edge and node addition on 4 coarsest levels are allowed), ``Musketeer Fine'' (30\% edge and node addition on 4 finest levels are allowed), and ``Musketeer All'' ( 15\% edge and node addition at all levels are allowed). The parameters are chosen such that the generated network has $3-4$ times the number of nodes and edges than the original network. We generated $30$ rescaled replicas for the same dataset as used in our previous experiment and compared the generated networks with the original network based on the following metrics: number of components, clustering coefficient, average degree, total degree-degree assortativity, average harmonic distance, modularity, pagerank and average betweenness centrality. 

\begin{figure}[t]
\includegraphics[width=1 \textwidth]{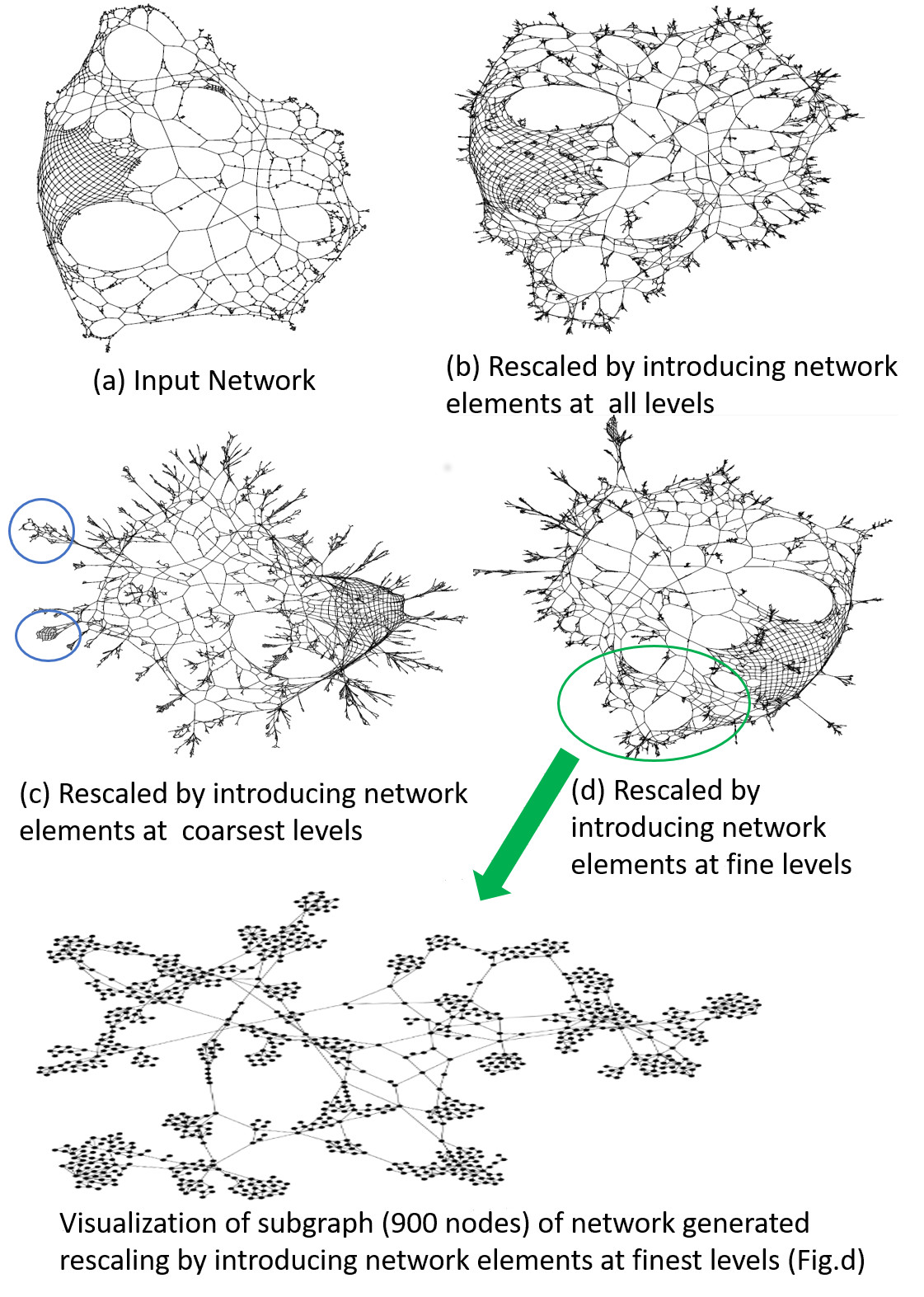}
\centering
\caption{Visualization of a road network from roadNet-TX with $2001$ nodes and $2957$ edges rescaled to at least $5900$ nodes and $7000$ edges. }\label{fig:Figure 16}
\end{figure}

\begin{figure}[t]
\includegraphics[width=1 \textwidth]{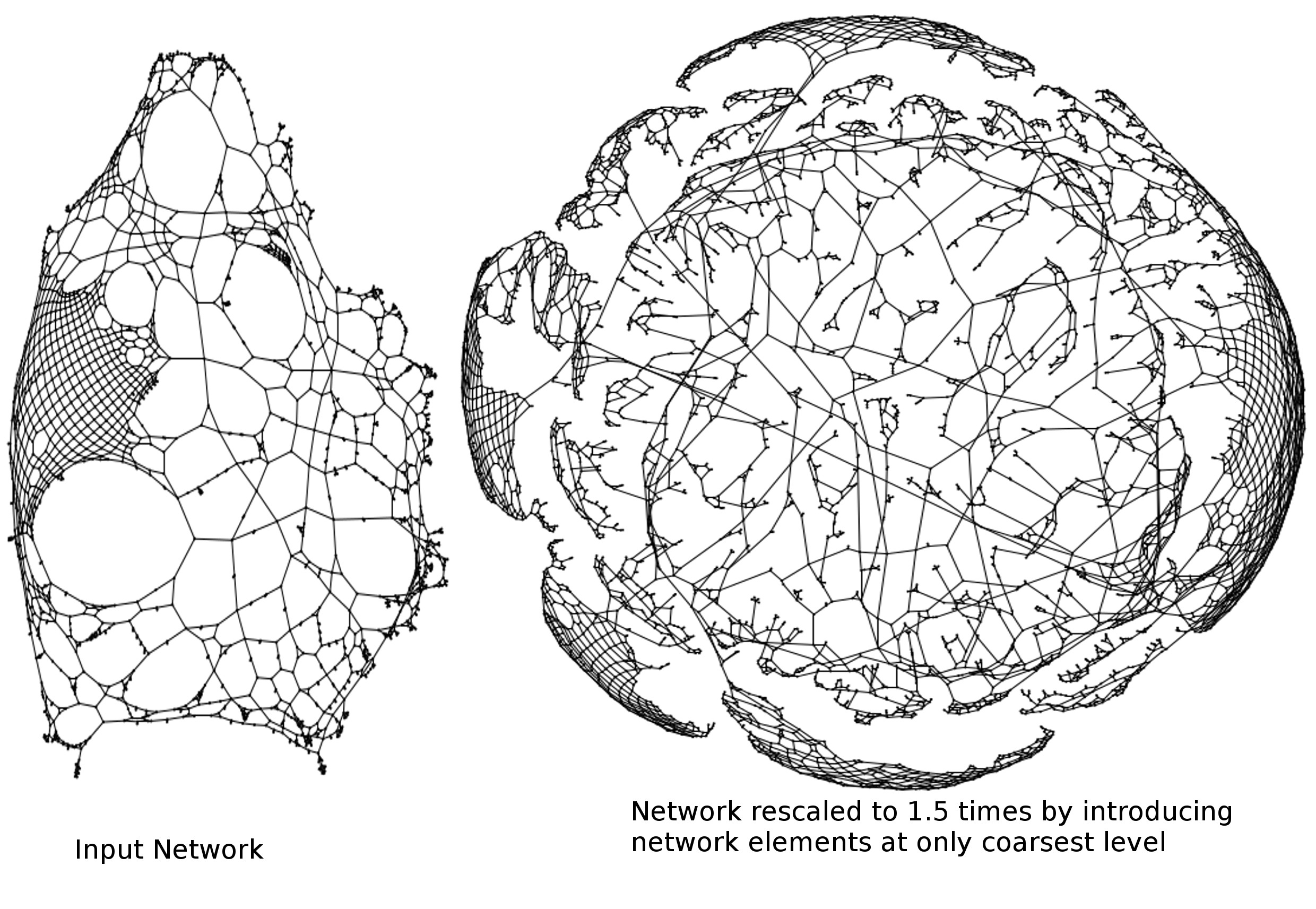}
\centering
\caption{Visualization of a road network from roadNet-TX with $2001$ nodes and $2957$ edges rescaled to at least $5028$ nodes and $7375$ edges. }\label{fig:Figure 17}
\end{figure}

\begin{figure}[t]
\includegraphics[width=1 \textwidth]{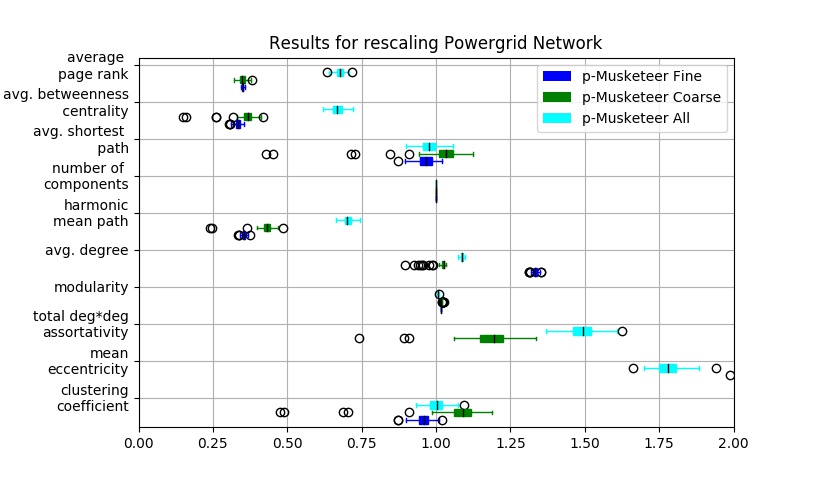}
\centering
\caption{Computational results for road network from roadNet-TX with $2001$ nodes and $2957$ edges rescaled to at least $6000$ nodes and $6500$ edges. }\label{fig:Figure 18}
\end{figure}

\begin{figure}[t]
\includegraphics[width=1 \textwidth]{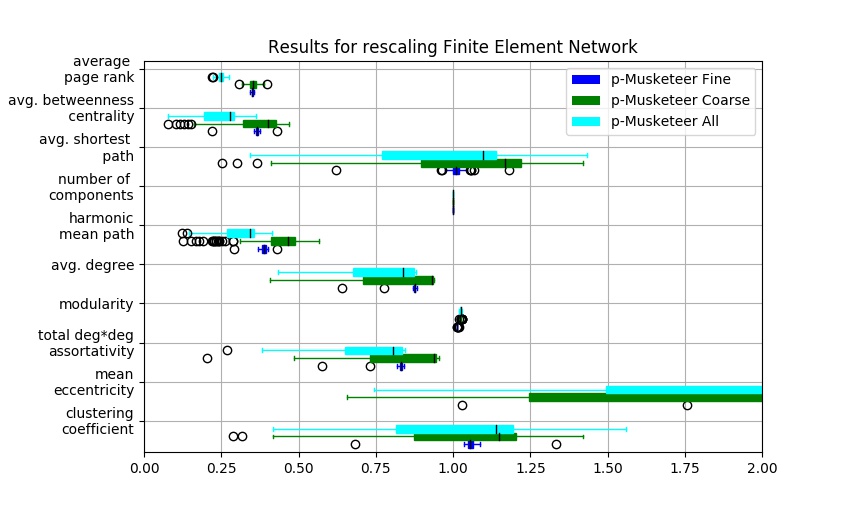}
\centering
\caption{Computational results for road network real water network with $407$ nodes and $459$ edges rescaled to at least $1098$ nodes and $1500$ edges.}\label{fig:Figure 19}
\end{figure}

\begin{figure}[t]
\includegraphics[width=1 \textwidth]{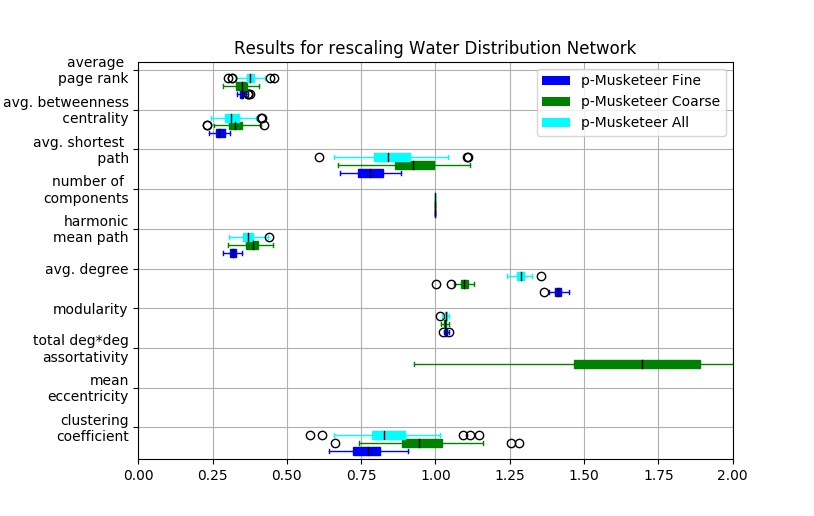}
\centering
\caption{Computational results for finite-element graph with $4704$ nodes and $13427$ edges rescaled to at least $12700$ nodes and $36000$ edges.}\label{fig:Figure 20}
\end{figure}

\begin{figure}[t]
\includegraphics[width=1 \textwidth]{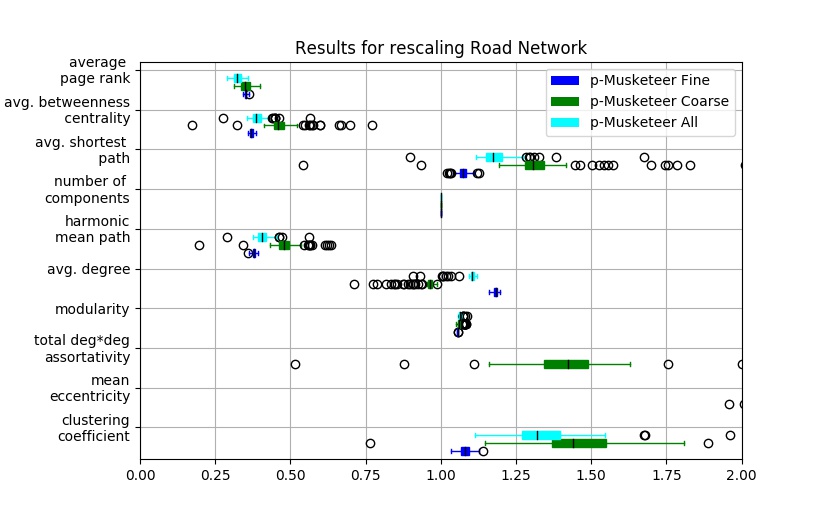}
\centering
\caption{Computational results for power grid graph opsahl-powergrid with with $4941$ nodes and $6211$ edges rescaled to at least $16500$ nodes and $27000$ edges.}\label{fig:Figure 21}
\end{figure}

\begin{figure}[t]
\includegraphics[width=1 \textwidth]{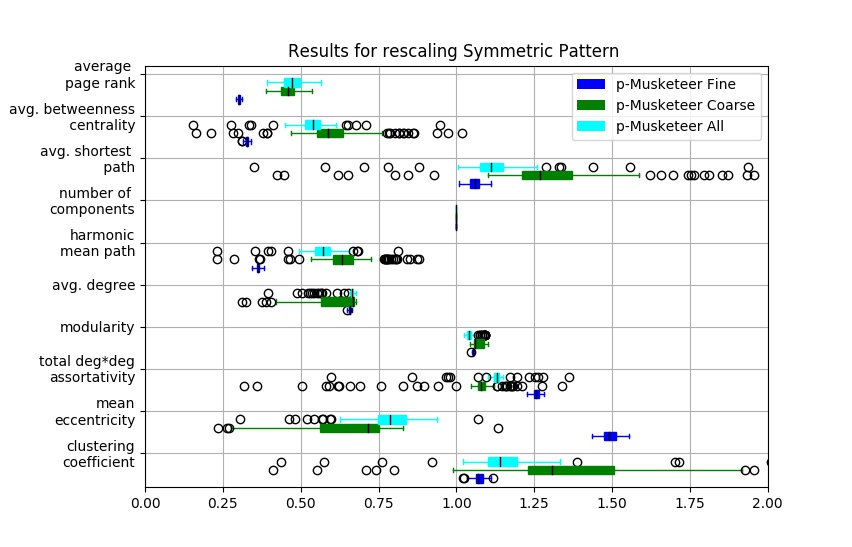}
\centering
\caption{Computational results on performance of planar Musketeer on symmetric pattern with $1141$ nodes and $3162$ edges to at least $2200$ nodes and $6000$ edges }\label{fig:Figure 22}
\end{figure}

\begin{figure}[t]
\includegraphics[width=1 \textwidth]{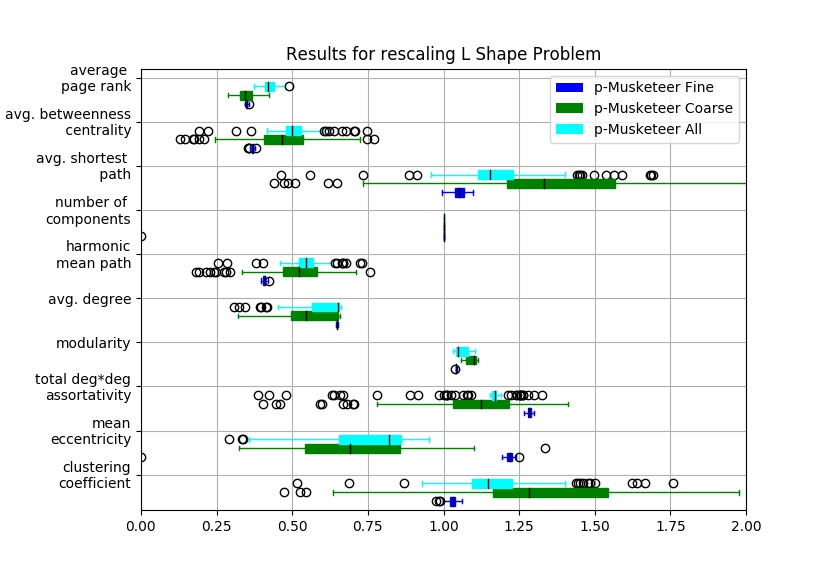}
\centering
\caption{Computational results on performance of planar Musketeer on graph for thermal L-Shape Problem with $3025$ nodes and $8904$ edges to at least $9000$ nodes and $12000$ edges }\label{fig:Figure 23}
\end{figure}

\begin{figure}[t]
\includegraphics[width=1 \textwidth]{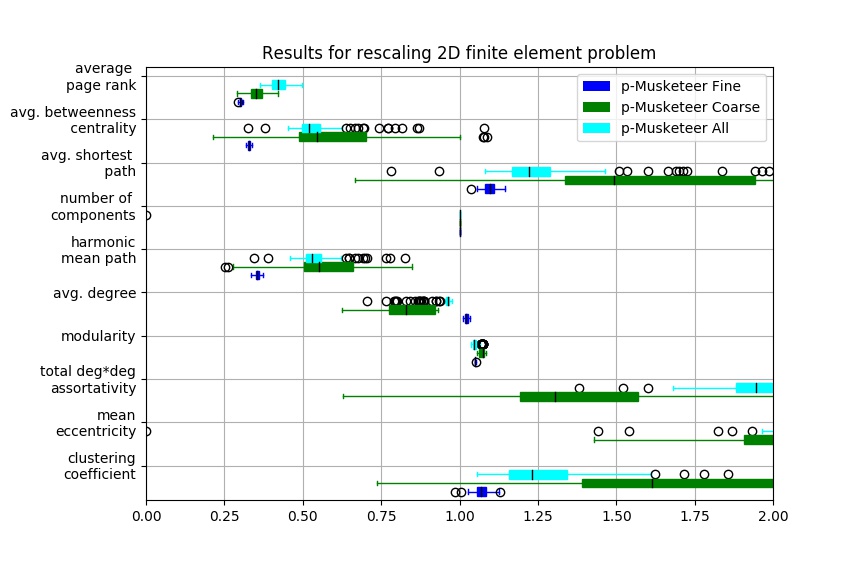}
\centering
\caption{Computational results on performance of planar Musketeer 2D finite element graph with $1866$ nodes and $3538$ edges rescaled to at least $4000$ nodes and $7000$ edges }\label{fig:Figure 24}
\end{figure}

The structural properties of the replicas were normalized such that $1$ denotes the property of original network. The comparison for $30$ experiments is presented in Figure \ref{fig:Figure 9}-\ref{fig:Figure 12}. As depicted in the plots we are able to preserve almost all the properties of original network even when the network is rescaled to more than $3$ times the original network. Also, there is no significant variance observed in properties for the three different sets of parameters (coarse,fine and all) used to generate rescaled networks. However, we observed that rescaling by introducing elements at finer levels results in high clustering coefficient in generated network. This is because the planarity constraint restricts addition of long edges (edges between nodes which are far from each other) which in turn forces the algorithm to connect new elements locally at each level $i$. In case the network elements are introduced at coarsest levels, the locally added edges and nodes are uncoarsened to several finer edges and nodes over the V-cycle of coarsening and uncoarsening, and the near neighbors at level $i$ are drifted apart at level $i+1$.However, network elements added at fine levels are not drifted as a result of levels of coarsening and uncoarsening as described above, and the edges still connect the nodes locally. Hence, we observe an increased number of triangles (Figure \ref{fig:Figure 16}) or high clustering coefficient (Figures \ref{fig:Figure 18} - \ref{fig:Figure 24}) for networks generated by introducing elements at fine level as compared to coarse level.
As depicted in Figure \ref{fig:Figure 16} when the network is rescaled by introducing new elements at only coarse levels, we find larger communities (e.g., mesh structures in case of our input road network) in the generated network, whereas if the network is rescaled at fine level we observe smaller communities. The amount of new introduced elements can be controlled by user input which is provided as node growth parameters at certain levels. The size of replicated and edited aggregated clusters in the rescaled network can be controlled by choosing larger node growth parameter for example in Fig. \ref{fig:Figure 17} we used node growth parameter as $1.5$ at the coarsest level to rescale the network to $1.5$ times. Depending on the application, one may want to add a postprocessing step which will create longer loops as in the original network of Fig. \ref{fig:Figure 17}. We created 13 links that close peripheral clusters by randomly choosing pairs of disconnected nodes, adding edge, and checking the planarity. However, although it may create a better visualization for the comparison with the original network, this step may not be desired by many application.

In our experiment we used $0.3$ as node growth rate for coarsest and finest level and $0.10$ when introducing network elements at all levels.


\section{Conclusions}
In this paper we introduced a multiscale planar graph generation framework and its implementation using Musketeer framework \cite{gutfraind2015multiscale} . Our evaluation suggest that multiscale planar graph generation method can generate realistic replicas of planar networks across domains with small loss of similarity. While there are clearly enough space for the improvement of this method, to the best of our knowledge, this is the first general purpose synthetic planar graph generation method that is able to produce realistic instances. 

Several future research directions can be explored. First, we would like to introduce the algebraic distance edge weighting scheme \cite{ChenS11} in order to more accurately preserve the distances during the uncoarsening. We have successfully used this improvement for network sparsification \cite{safro:spars} and several combinatorial optimization solvers \cite{amg-sss12,safro:relaxml}. Second, it would be interesting to investigate whether one should control the size of aggregates to preserve uniform coarsening, a multilevel technique that is well known in graph partitioning \cite{amg-sss12,buluc2016recent}. The role of uniform coarsening is not well understood in multiscale graph generation.

There exist several further research directions that are application dependent. For example, approaches for assessing how much variation is desirable in the generation and how to decide whether enough changes have been introduced can vary from application to application. However, a unified strategy to address this issue would be very helpful.

	\bibliographystyle{plain}

\begin{thebibliography}{52}
\ifx \bisbn   \undefined \def \bisbn  #1{ISBN #1}\fi
\ifx \binits  \undefined \def \binits#1{#1}\fi
\ifx \bauthor  \undefined \def \bauthor#1{#1}\fi
\ifx \batitle  \undefined \def \batitle#1{#1}\fi
\ifx \bjtitle  \undefined \def \bjtitle#1{#1}\fi
\ifx \bvolume  \undefined \def \bvolume#1{\textbf{#1}}\fi
\ifx \byear  \undefined \def \byear#1{#1}\fi
\ifx \bissue  \undefined \def \bissue#1{#1}\fi
\ifx \bfpage  \undefined \def \bfpage#1{#1}\fi
\ifx \blpage  \undefined \def \blpage #1{#1}\fi
\ifx \burl  \undefined \def \burl#1{\textsf{#1}}\fi
\ifx \doiurl  \undefined \def \doiurl#1{\textsf{#1}}\fi
\ifx \betal  \undefined \def \betal{\textit{et al.}}\fi
\ifx \binstitute  \undefined \def \binstitute#1{#1}\fi
\ifx \binstitutionaled  \undefined \def \binstitutionaled#1{#1}\fi
\ifx \bctitle  \undefined \def \bctitle#1{#1}\fi
\ifx \beditor  \undefined \def \beditor#1{#1}\fi
\ifx \bpublisher  \undefined \def \bpublisher#1{#1}\fi
\ifx \bbtitle  \undefined \def \bbtitle#1{#1}\fi
\ifx \bedition  \undefined \def \bedition#1{#1}\fi
\ifx \bseriesno  \undefined \def \bseriesno#1{#1}\fi
\ifx \blocation  \undefined \def \blocation#1{#1}\fi
\ifx \bsertitle  \undefined \def \bsertitle#1{#1}\fi
\ifx \bsnm \undefined \def \bsnm#1{#1}\fi
\ifx \bsuffix \undefined \def \bsuffix#1{#1}\fi
\ifx \bparticle \undefined \def \bparticle#1{#1}\fi
\ifx \barticle \undefined \def \barticle#1{#1}\fi
\ifx \bconfdate \undefined \def \bconfdate #1{#1}\fi
\ifx \botherref \undefined \def \botherref #1{#1}\fi
\ifx \url \undefined \def \url#1{\textsf{#1}}\fi
\ifx \bchapter \undefined \def \bchapter#1{#1}\fi
\ifx \bbook \undefined \def \bbook#1{#1}\fi
\ifx \bcomment \undefined \def \bcomment#1{#1}\fi
\ifx \oauthor \undefined \def \oauthor#1{#1}\fi
\ifx \citeauthoryear \undefined \def \citeauthoryear#1{#1}\fi
\ifx \endbibitem  \undefined \def \endbibitem {}\fi
\ifx \bconflocation  \undefined \def \bconflocation#1{#1}\fi
\ifx \arxivurl  \undefined \def \arxivurl#1{\textsf{#1}}\fi
\csname PreBibitemsHook\endcsname

\bibitem{newman2018networks}
\begin{bbook}
\bauthor{\bsnm{Newman}, \binits{M.}}:
\bbtitle{Networks}.
\bpublisher{Oxford university press}, \blocation{???}
(\byear{2018})
\end{bbook}
\endbibitem

\bibitem{gutfraind2015multiscale}
\begin{bchapter}
\bauthor{\bsnm{Gutfraind}, \binits{A.}},
\bauthor{\bsnm{Safro}, \binits{I.}},
\bauthor{\bsnm{Meyers}, \binits{L.A.}}:
\bctitle{Multiscale network generation}.
In: \bbtitle{18th IEEE International Conference on Information Fusion
  (Fusion)},
pp. \bfpage{158}--\blpage{165}
(\byear{2015}).
\bcomment{IEEE}
\end{bchapter}
\endbibitem

\bibitem{staudt17generating-ans}
\begin{barticle}
\bauthor{\bsnm{Staudt}, \binits{C.L.}},
\bauthor{\bsnm{Hamann}, \binits{M.}},
\bauthor{\bsnm{Gutfraind}, \binits{A.}},
\bauthor{\bsnm{Safro}, \binits{I.}},
\bauthor{\bsnm{Meyerhenke}, \binits{H.}}:
\batitle{Generating realistic scaled complex networks}.
\bjtitle{Applied Network Science}
\bvolume{2}(\bissue{1}),
\bfpage{36}
(\byear{2017}).
doi:\doiurl{10.1007/s41109-017-0054-z}
\end{barticle}
\endbibitem

\bibitem{meinert2011experimental}
\begin{bchapter}
\bauthor{\bsnm{Meinert}, \binits{S.}},
\bauthor{\bsnm{Wagner}, \binits{D.}}:
\bctitle{An experimental study on generating planar graphs}.
In: \bbtitle{Frontiers in Algorithmics and Algorithmic Aspects in Information
  and Management},
pp. \bfpage{375}--\blpage{387}.
\bpublisher{Springer}, \blocation{???}
(\byear{2011})
\end{bchapter}
\endbibitem

\bibitem{gutfraind2012multiscale}
\begin{botherref}
\oauthor{\bsnm{Gutfraind}, \binits{A.}},
\oauthor{\bsnm{Meyers}, \binits{L.A.}},
\oauthor{\bsnm{Safro}, \binits{I.}}:
Multiscale network generation.
CoRR
\textbf{abs/1207.4266}
(2012).
\arxivurl{1207.4266}
\end{botherref}
\endbibitem

\bibitem{barthelemy2011spatial}
\begin{barticle}
\bauthor{\bsnm{Barth{\'e}lemy}, \binits{M.}}:
\batitle{Spatial networks}.
\bjtitle{Physics Reports}
\bvolume{499}(\bissue{1-3}),
\bfpage{1}--\blpage{101}
(\byear{2011})
\end{barticle}
\endbibitem

\bibitem{hunter2008ergm}
\begin{barticle}
\bauthor{\bsnm{Hunter}, \binits{D.R.}},
\bauthor{\bsnm{Handcock}, \binits{M.S.}},
\bauthor{\bsnm{Butts}, \binits{C.T.}},
\bauthor{\bsnm{Goodreau}, \binits{S.M.}},
\bauthor{\bsnm{Morris}, \binits{M.}}:
\batitle{ergm: A package to fit, simulate and diagnose exponential-family
  models for networks}.
\bjtitle{Journal of statistical software}
\bvolume{24}(\bissue{3}),
\bfpage{54860}
(\byear{2008})
\end{barticle}
\endbibitem

\bibitem{aiello2000random}
\begin{bchapter}
\bauthor{\bsnm{Aiello}, \binits{W.}},
\bauthor{\bsnm{Chung}, \binits{F.}},
\bauthor{\bsnm{Lu}, \binits{L.}}:
\bctitle{A random graph model for massive graphs}.
In: \bbtitle{Proceedings of the Thirty-second Annual ACM Symposium on Theory of
  Computing},
pp. \bfpage{171}--\blpage{180}
(\byear{2000}).
\bcomment{Acm}
\end{bchapter}
\endbibitem

\bibitem{karrer2011stochastic}
\begin{barticle}
\bauthor{\bsnm{Karrer}, \binits{B.}},
\bauthor{\bsnm{Newman}, \binits{M.E.}}:
\batitle{Stochastic blockmodels and community structure in networks}.
\bjtitle{Physical review E}
\bvolume{83}(\bissue{1}),
\bfpage{016107}
(\byear{2011})
\end{barticle}
\endbibitem

\bibitem{fronczak2013exponential}
\begin{barticle}
\bauthor{\bsnm{Fronczak}, \binits{P.}},
\bauthor{\bsnm{Fronczak}, \binits{A.}},
\bauthor{\bsnm{Bujok}, \binits{M.}}:
\batitle{Exponential random graph models for networks with community
  structure}.
\bjtitle{Physical Review E}
\bvolume{88}(\bissue{3}),
\bfpage{032810}
(\byear{2013})
\end{barticle}
\endbibitem

\bibitem{van2019reconstructing}
\begin{botherref}
\oauthor{\bparticle{van Lidth~de} \bsnm{Jeude}, \binits{J.}},
\oauthor{\bsnm{Di~Clemente}, \binits{R.}},
\oauthor{\bsnm{Caldarelli}, \binits{G.}},
\oauthor{\bsnm{Saracco}, \binits{F.}},
\oauthor{\bsnm{Squartini}, \binits{T.}}:
Reconstructing mesoscale network structures.
Complexity
\textbf{2019}
(2019)
\end{botherref}
\endbibitem

\bibitem{seshadhri2012community}
\begin{barticle}
\bauthor{\bsnm{Seshadhri}, \binits{C.}},
\bauthor{\bsnm{Kolda}, \binits{T.G.}},
\bauthor{\bsnm{Pinar}, \binits{A.}}:
\batitle{Community structure and scale-free collections of
  erd{\H{o}}s-r{\'e}nyi graphs}.
\bjtitle{Physical Review E}
\bvolume{85}(\bissue{5}),
\bfpage{056109}
(\byear{2012})
\end{barticle}
\endbibitem

\bibitem{aiello2001random}
\begin{barticle}
\bauthor{\bsnm{Aiello}, \binits{W.}},
\bauthor{\bsnm{Chung}, \binits{F.}},
\bauthor{\bsnm{Lu}, \binits{L.}}:
\batitle{A random graph model for power law graphs}.
\bjtitle{Experimental Mathematics}
\bvolume{10}(\bissue{1}),
\bfpage{53}--\blpage{66}
(\byear{2001})
\end{barticle}
\endbibitem

\bibitem{chakrabarti2004r}
\begin{bchapter}
\bauthor{\bsnm{Chakrabarti}, \binits{D.}},
\bauthor{\bsnm{Zhan}, \binits{Y.}},
\bauthor{\bsnm{Faloutsos}, \binits{C.}}:
\bctitle{R-mat: A recursive model for graph mining}.
In: \bbtitle{Proceedings of the 2004 SIAM International Conference on Data
  Mining},
pp. \bfpage{442}--\blpage{446}
(\byear{2004}).
\bcomment{SIAM}
\end{bchapter}
\endbibitem

\bibitem{mahdian2007stochastic}
\begin{bchapter}
\bauthor{\bsnm{Mahdian}, \binits{M.}},
\bauthor{\bsnm{Xu}, \binits{Y.}}:
\bctitle{Stochastic kronecker graphs}.
In: \bbtitle{International Workshop on Algorithms and Models for the
  Web-Graph},
pp. \bfpage{179}--\blpage{186}
(\byear{2007}).
\bcomment{Springer}
\end{bchapter}
\endbibitem

\bibitem{palla2010multifractal}
\begin{barticle}
\bauthor{\bsnm{Palla}, \binits{G.}},
\bauthor{\bsnm{Lov{\'a}sz}, \binits{L.}},
\bauthor{\bsnm{Vicsek}, \binits{T.}}:
\batitle{Multifractal network generator}.
\bjtitle{Proceedings of the National Academy of Sciences}
\bvolume{107}(\bissue{17}),
\bfpage{7640}
(\byear{2010})
\end{barticle}
\endbibitem

\bibitem{Newman2010}
\begin{bbook}
\bauthor{\bsnm{Newman}, \binits{M.}}:
\bbtitle{Networks: An Introduction}.
\bpublisher{Oxford University Press, Inc.},
\blocation{New York, NY, USA}
(\byear{2010})
\end{bbook}
\endbibitem

\bibitem{Tabourier:2011:GCR:1963190.2063515}
\begin{barticle}
\bauthor{\bsnm{Tabourier}, \binits{L.}},
\bauthor{\bsnm{Roth}, \binits{C.}},
\bauthor{\bsnm{Cointet}, \binits{J.-P.}}:
\batitle{Generating constrained random graphs using multiple edge switches}.
\bjtitle{J. Exp. Algorithmics}
\bvolume{16},
\bfpage{1}--\blpage{71117115}
(\byear{2011}).
doi:\doiurl{10.1145/1963190.2063515}
\end{barticle}
\endbibitem

\bibitem{rao1996markov}
\begin{botherref}
\oauthor{\bsnm{Rao}, \binits{A.R.}},
\oauthor{\bsnm{Jana}, \binits{R.}},
\oauthor{\bsnm{Bandyopadhyay}, \binits{S.}}:
A markov chain monte carlo method for generating random (0, 1)-matrices with
  given marginals.
Sankhy{\=a}: The Indian Journal of Statistics, Series A,
225--242
(1996)
\end{botherref}
\endbibitem

\bibitem{safro:relaxml}
\begin{barticle}
\bauthor{\bsnm{Ron}, \binits{D.}},
\bauthor{\bsnm{Safro}, \binits{I.}},
\bauthor{\bsnm{Brandt}, \binits{A.}}:
\batitle{Relaxation-based coarsening and multiscale graph organization}.
\bjtitle{Multiscale Modeling \& Simulation}
\bvolume{9}(\bissue{1}),
\bfpage{407}--\blpage{423}
(\byear{2011})
\end{barticle}
\endbibitem

\bibitem{safro:mlvsp}
\begin{barticle}
\bauthor{\bsnm{Hager}, \binits{W.W.}},
\bauthor{\bsnm{Hungerford}, \binits{J.T.}},
\bauthor{\bsnm{Safro}, \binits{I.}}:
\batitle{A multilevel bilinear programming algorithm for the vertex separator
  problem}.
\bjtitle{Computational Optimization and Applications}
\bvolume{69}(\bissue{1}),
\bfpage{189}--\blpage{223}
(\byear{2018})
\end{barticle}
\endbibitem

\bibitem{SafroRB08}
\begin{botherref}
\oauthor{\bsnm{Safro}, \binits{I.}},
\oauthor{\bsnm{Ron}, \binits{D.}},
\oauthor{\bsnm{Brandt}, \binits{A.}}:
Multilevel algorithms for linear ordering problems.
ACM Journal of Experimental Algorithmics
\textbf{13}
(2008)
\end{botherref}
\endbibitem

\bibitem{safro2006graph}
\begin{barticle}
\bauthor{\bsnm{Safro}, \binits{I.}},
\bauthor{\bsnm{Ron}, \binits{D.}},
\bauthor{\bsnm{Brandt}, \binits{A.}}:
\batitle{Graph minimum linear arrangement by multilevel weighted edge
  contractions}.
\bjtitle{Journal of Algorithms}
\bvolume{60}(\bissue{1}),
\bfpage{24}--\blpage{41}
(\byear{2006})
\end{barticle}
\endbibitem

\bibitem{SafroT11}
\begin{barticle}
\bauthor{\bsnm{Safro}, \binits{I.}},
\bauthor{\bsnm{Temkin}, \binits{B.}}:
\batitle{Multiscale approach for the network compression-friendly ordering}.
\bjtitle{J. Discrete Algorithms}
\bvolume{9}(\bissue{2}),
\bfpage{190}--\blpage{202}
(\byear{2011})
\end{barticle}
\endbibitem

\bibitem{staudt2014networkit}
\begin{botherref}
\oauthor{\bsnm{Staudt}, \binits{C.}},
\oauthor{\bsnm{Sazonovs}, \binits{A.}},
\oauthor{\bsnm{Meyerhenke}, \binits{H.}}:
Networkit: An interactive tool suite for high-performance network analysis.
CoRR, abs/1403.3005
(2014)
\end{botherref}
\endbibitem

\bibitem{tutte1963census}
\begin{barticle}
\bauthor{\bsnm{Tutte}, \binits{W.T.}}:
\batitle{A census of planar maps}.
\bjtitle{Canad. J. Math}
\bvolume{15}(\bissue{2}),
\bfpage{249}--\blpage{271}
(\byear{1963})
\end{barticle}
\endbibitem

\bibitem{brinkmann2007fast}
\begin{barticle}
\bauthor{\bsnm{Brinkmann}, \binits{G.}},
\bauthor{\bsnm{McKay}, \binits{B.D.}}, \betal:
\batitle{Fast generation of planar graphs}.
\bjtitle{MATCH Commun. Math. Comput. Chem}
\bvolume{58}(\bissue{2}),
\bfpage{323}--\blpage{357}
(\byear{2007})
\end{barticle}
\endbibitem

\bibitem{brinkmann2011program}
\begin{botherref}
\oauthor{\bsnm{Brinkmann}, \binits{G.}}:
Program fullgen-a program for generating nonisomorphic fullerenes.
see http://cs. anu. edu. au/bdm/plantri
(2011)
\end{botherref}
\endbibitem

\bibitem{denise1996random}
\begin{botherref}
\oauthor{\bsnm{Denise}, \binits{A.}},
\oauthor{\bsnm{Vasconcellos}, \binits{M.}},
\oauthor{\bsnm{Welsh}, \binits{D.J.}}:
The random planar graph.
Congressus numerantium,
61--80
(1996)
\end{botherref}
\endbibitem

\bibitem{shewchuk1996triangle}
\begin{bchapter}
\bauthor{\bsnm{Shewchuk}, \binits{J.R.}}:
\bctitle{Triangle: Engineering a 2d quality mesh generator and delaunay
  triangulator}.
In: \bbtitle{Applied Computational Geometry Towards Geometric Engineering},
pp. \bfpage{203}--\blpage{222}.
\bpublisher{Springer}, \blocation{???}
(\byear{1996})
\end{bchapter}
\endbibitem

\bibitem{ruppert1995delaunay}
\begin{barticle}
\bauthor{\bsnm{Ruppert}, \binits{J.}}:
\batitle{A delaunay refinement algorithm for quality 2-dimensional mesh
  generation}.
\bjtitle{Journal of algorithms}
\bvolume{18}(\bissue{3}),
\bfpage{548}--\blpage{585}
(\byear{1995})
\end{barticle}
\endbibitem

\bibitem{gilbert1961random}
\begin{barticle}
\bauthor{\bsnm{Gilbert}, \binits{E.N.}}:
\batitle{Random plane networks}.
\bjtitle{Journal of the Society for Industrial and Applied Mathematics}
\bvolume{9}(\bissue{4}),
\bfpage{533}--\blpage{543}
(\byear{1961})
\end{barticle}
\endbibitem

\bibitem{erdds1959random}
\begin{barticle}
\bauthor{\bsnm{Erd\"{o}s}, \binits{P.}},
\bauthor{\bsnm{R\'{e}nyi}, \binits{A.}}:
\batitle{{On random graphs, I}}.
\bjtitle{Publicationes Mathematicae (Debrecen)}
\bvolume{6},
\bfpage{290}--\blpage{297}
(\byear{1959})
\end{barticle}
\endbibitem

\bibitem{gerke2004number}
\begin{barticle}
\bauthor{\bsnm{Gerke}, \binits{S.}},
\bauthor{\bsnm{McDiarmid}, \binits{C.}}:
\batitle{On the number of edges in random planar graphs}.
\bjtitle{Combinatorics, Probability and Computing}
\bvolume{13}(\bissue{2}),
\bfpage{165}--\blpage{183}
(\byear{2004})
\end{barticle}
\endbibitem

\bibitem{mcdiarmid2005random}
\begin{barticle}
\bauthor{\bsnm{McDiarmid}, \binits{C.}},
\bauthor{\bsnm{Steger}, \binits{A.}},
\bauthor{\bsnm{Welsh}, \binits{D.J.}}:
\batitle{Random planar graphs}.
\bjtitle{Journal of Combinatorial Theory, Series B}
\bvolume{93}(\bissue{2}),
\bfpage{187}--\blpage{205}
(\byear{2005})
\end{barticle}
\endbibitem

\bibitem{cura2015streetgen}
\begin{barticle}
\bauthor{\bsnm{Cura}, \binits{R.}},
\bauthor{\bsnm{Perret}, \binits{J.}},
\bauthor{\bsnm{Paparoditis}, \binits{N.}}:
\batitle{Streetgen: In-base procedural-based road generation}.
\bjtitle{ISPRS Annals of the Photogrammetry, Remote Sensing and Spatial
  Information Sciences}
\bvolume{2},
\bfpage{409}
(\byear{2015})
\end{barticle}
\endbibitem

\bibitem{wang2008generating}
\begin{bchapter}
\bauthor{\bsnm{Wang}, \binits{Z.}},
\bauthor{\bsnm{Thomas}, \binits{R.J.}},
\bauthor{\bsnm{Scaglione}, \binits{A.}}:
\bctitle{Generating random topology power grids}.
In: \bbtitle{Hawaii International Conference on System Sciences, Proceedings of
  the 41st Annual},
pp. \bfpage{183}--\blpage{183}
(\byear{2008}).
\bcomment{IEEE}
\end{bchapter}
\endbibitem

\bibitem{sitzenfrei2013automatic}
\begin{barticle}
\bauthor{\bsnm{Sitzenfrei}, \binits{R.}},
\bauthor{\bsnm{M{\"o}derl}, \binits{M.}},
\bauthor{\bsnm{Rauch}, \binits{W.}}:
\batitle{Automatic generation of water distribution systems based on gis data}.
\bjtitle{Environmental modelling \& software}
\bvolume{47},
\bfpage{138}--\blpage{147}
(\byear{2013})
\end{barticle}
\endbibitem

\bibitem{muranho2012waternetgen}
\begin{barticle}
\bauthor{\bsnm{Muranho}, \binits{J.}},
\bauthor{\bsnm{Ferreira}, \binits{A.}},
\bauthor{\bsnm{Sousa}, \binits{J.}},
\bauthor{\bsnm{Gomes}, \binits{A.}},
\bauthor{\bsnm{Marques}, \binits{A.S.}}:
\batitle{Waternetgen: an epanet extension for automatic water distribution
  network models generation and pipe sizing}.
\bjtitle{Water science and technology: water supply}
\bvolume{12}(\bissue{1}),
\bfpage{117}--\blpage{123}
(\byear{2012})
\end{barticle}
\endbibitem

\bibitem{rossman2000epanet}
\begin{botherref}
\oauthor{\bsnm{Rossman}, \binits{L.A.}}, et al.:
Epanet 2: users manual
(2000)
\end{botherref}
\endbibitem

\bibitem{staudt2016generating}
\begin{bchapter}
\bauthor{\bsnm{Staudt}, \binits{C.L.}},
\bauthor{\bsnm{Hamann}, \binits{M.}},
\bauthor{\bsnm{Safro}, \binits{I.}},
\bauthor{\bsnm{Gutfraind}, \binits{A.}},
\bauthor{\bsnm{Meyerhenke}, \binits{H.}}:
\bctitle{Generating scaled replicas of real-world complex networks}.
In: \bbtitle{International Workshop on Complex Networks and Their
  Applications},
pp. \bfpage{17}--\blpage{28}
(\byear{2016}).
\bcomment{Springer}
\end{bchapter}
\endbibitem

\bibitem{vlsicad}
\begin{bchapter}
\bauthor{\bsnm{Brandt}, \binits{A.}},
\bauthor{\bsnm{Ron}, \binits{D.}}:
\bctitle{Chapter 1 : Multigrid solvers and multilevel optimization strategies}.
In: \beditor{\bsnm{Cong}, \binits{J.}},
\beditor{\bsnm{Shinnerl}, \binits{J.R.}} (eds.)
\bbtitle{Multilevel Optimization and VLSICAD}.
\bpublisher{Kluwer}, \blocation{???}
(\byear{2003})
\end{bchapter}
\endbibitem

\bibitem{thomassen1981kuratowski}
\begin{barticle}
\bauthor{\bsnm{Thomassen}, \binits{C.}}:
\batitle{Kuratowski's theorem}.
\bjtitle{Journal of Graph Theory}
\bvolume{5}(\bissue{3}),
\bfpage{225}--\blpage{241}
(\byear{1981})
\end{barticle}
\endbibitem

\bibitem{ostfeld2008battle}
\begin{barticle}
\bauthor{\bsnm{Ostfeld}, \binits{A.}},
\bauthor{\bsnm{Uber}, \binits{J.G.}},
\bauthor{\bsnm{Salomons}, \binits{E.}},
\bauthor{\bsnm{Berry}, \binits{J.W.}},
\bauthor{\bsnm{Hart}, \binits{W.E.}},
\bauthor{\bsnm{Phillips}, \binits{C.A.}},
\bauthor{\bsnm{Watson}, \binits{J.-P.}},
\bauthor{\bsnm{Dorini}, \binits{G.}},
\bauthor{\bsnm{Jonkergouw}, \binits{P.}},
\bauthor{\bsnm{Kapelan}, \binits{Z.}}, \betal:
\batitle{The battle of the water sensor networks (bwsn): A design challenge for
  engineers and algorithms}.
\bjtitle{Journal of Water Resources Planning and Management}
\bvolume{134}(\bissue{6}),
\bfpage{556}--\blpage{568}
(\byear{2008})
\end{barticle}
\endbibitem

\bibitem{leskovec2009community}
\begin{barticle}
\bauthor{\bsnm{Leskovec}, \binits{J.}},
\bauthor{\bsnm{Lang}, \binits{K.J.}},
\bauthor{\bsnm{Dasgupta}, \binits{A.}},
\bauthor{\bsnm{Mahoney}, \binits{M.W.}}:
\batitle{Community structure in large networks: Natural cluster sizes and the
  absence of large well-defined clusters}.
\bjtitle{Internet Mathematics}
\bvolume{6}(\bissue{1}),
\bfpage{29}--\blpage{123}
(\byear{2009})
\end{barticle}
\endbibitem

\bibitem{snapnets}
\begin{botherref}
\oauthor{\bsnm{Leskovec}, \binits{J.}},
\oauthor{\bsnm{Krevl}, \binits{A.}}:
{SNAP Datasets}: {Stanford} Large Network Dataset Collection.
\url{http://snap.stanford.edu/data}
(2014)
\end{botherref}
\endbibitem

\bibitem{Davis1997}
\begin{botherref}
\oauthor{\bsnm{Davis}, \binits{T.}}:
University of {F}lorida {S}parse {M}atrix {C}ollection.
NA Digest
\textbf{97}(23)
(1997)
\end{botherref}
\endbibitem

\bibitem{chimani2013open}
\begin{barticle}
\bauthor{\bsnm{Chimani}, \binits{M.}},
\bauthor{\bsnm{Gutwenger}, \binits{C.}},
\bauthor{\bsnm{J{\"u}nger}, \binits{M.}},
\bauthor{\bsnm{Klau}, \binits{G.W.}},
\bauthor{\bsnm{Klein}, \binits{K.}},
\bauthor{\bsnm{Mutzel}, \binits{P.}}:
\batitle{The open graph drawing framework (ogdf).}
\bjtitle{Handbook of Graph Drawing and Visualization}
\bvolume{2011},
\bfpage{543}--\blpage{569}
(\byear{2013})
\end{barticle}
\endbibitem

\bibitem{ChenS11}
\begin{barticle}
\bauthor{\bsnm{Chen}, \binits{J.}},
\bauthor{\bsnm{Safro}, \binits{I.}}:
\batitle{Algebraic distance on graphs}.
\bjtitle{SIAM J. Scientific Computing}
\bvolume{33}(\bissue{6}),
\bfpage{3468}--\blpage{3490}
(\byear{2011})
\end{barticle}
\endbibitem

\bibitem{safro:spars}
\begin{barticle}
\bauthor{\bsnm{John}, \binits{E.}},
\bauthor{\bsnm{Safro}, \binits{I.}}:
\batitle{Single-and multi-level network sparsification by algebraic distance}.
\bjtitle{Journal of Complex Networks}
\bvolume{5}(\bissue{3}),
\bfpage{352}--\blpage{388}
(\byear{2016})
\end{barticle}
\endbibitem

\bibitem{amg-sss12}
\begin{barticle}
\bauthor{\bsnm{Safro}, \binits{I.}},
\bauthor{\bsnm{Sanders}, \binits{P.}},
\bauthor{\bsnm{Schulz}, \binits{C.}}:
\batitle{Advanced coarsening schemes for graph partitioning}.
\bjtitle{ACM Journal of Experimental Algorithmics (JEA)}
\bvolume{19},
\bfpage{2}--\blpage{2}
(\byear{2015})
\end{barticle}
\endbibitem

\bibitem{buluc2016recent}
\begin{bchapter}
\bauthor{\bsnm{Bulu{\c{c}}}, \binits{A.}},
\bauthor{\bsnm{Meyerhenke}, \binits{H.}},
\bauthor{\bsnm{Safro}, \binits{I.}},
\bauthor{\bsnm{Sanders}, \binits{P.}},
\bauthor{\bsnm{Schulz}, \binits{C.}}:
\bctitle{Recent advances in graph partitioning}.
In: \bbtitle{Algorithm Engineering: Selected Results and Surveys},
pp. \bfpage{117}--\blpage{158}.
\bpublisher{Springer}, \blocation{???}
(\byear{2016})
\end{bchapter}
\endbibitem

\end{thebibliography}

\newcommand{\BMCxmlcomment}[1]{}

\BMCxmlcomment{

<refgrp>

<bibl id="B1">
  <title><p>Networks</p></title>
  <aug>
    <au><snm>Newman</snm><fnm>M</fnm></au>
  </aug>
  <publisher>Oxford university press</publisher>
  <pubdate>2018</pubdate>
</bibl>

<bibl id="B2">
  <title><p>Multiscale network generation</p></title>
  <aug>
    <au><snm>Gutfraind</snm><fnm>A</fnm></au>
    <au><snm>Safro</snm><fnm>I</fnm></au>
    <au><snm>Meyers</snm><fnm>LA</fnm></au>
  </aug>
  <source>18th IEEE International Conference on Information Fusion
  (Fusion)</source>
  <pubdate>2015</pubdate>
  <fpage>158</fpage>
  <lpage>-165</lpage>
</bibl>

<bibl id="B3">
  <title><p>Generating realistic scaled complex networks</p></title>
  <aug>
    <au><snm>Staudt</snm><fnm>CL</fnm></au>
    <au><snm>Hamann</snm><fnm>M</fnm></au>
    <au><snm>Gutfraind</snm><fnm>A</fnm></au>
    <au><snm>Safro</snm><fnm>I</fnm></au>
    <au><snm>Meyerhenke</snm><fnm>H</fnm></au>
  </aug>
  <source>Applied Network Science</source>
  <pubdate>2017</pubdate>
  <volume>2</volume>
  <issue>1</issue>
  <fpage>36</fpage>
  <url>https://doi.org/10.1007/s41109-017-0054-z</url>
</bibl>

<bibl id="B4">
  <title><p>An experimental study on generating planar graphs</p></title>
  <aug>
    <au><snm>Meinert</snm><fnm>S</fnm></au>
    <au><snm>Wagner</snm><fnm>D</fnm></au>
  </aug>
  <source>Frontiers in Algorithmics and Algorithmic Aspects in Information and
  Management</source>
  <publisher>Springer</publisher>
  <pubdate>2011</pubdate>
  <fpage>375</fpage>
  <lpage>-387</lpage>
</bibl>

<bibl id="B5">
  <title><p>Multiscale Network Generation</p></title>
  <aug>
    <au><snm>Gutfraind</snm><fnm>A</fnm></au>
    <au><snm>Meyers</snm><fnm>LA</fnm></au>
    <au><snm>Safro</snm><fnm>I</fnm></au>
  </aug>
  <source>CoRR</source>
  <pubdate>2012</pubdate>
  <volume>abs/1207.4266</volume>
  <url>http://arxiv.org/abs/1207.4266</url>
</bibl>

<bibl id="B6">
  <title><p>Spatial networks</p></title>
  <aug>
    <au><snm>Barth{\'e}lemy</snm><fnm>M</fnm></au>
  </aug>
  <source>Physics Reports</source>
  <publisher>Elsevier</publisher>
  <pubdate>2011</pubdate>
  <volume>499</volume>
  <issue>1-3</issue>
  <fpage>1</fpage>
  <lpage>-101</lpage>
</bibl>

<bibl id="B7">
  <title><p>ergm: A package to fit, simulate and diagnose exponential-family
  models for networks</p></title>
  <aug>
    <au><snm>Hunter</snm><fnm>DR</fnm></au>
    <au><snm>Handcock</snm><fnm>MS</fnm></au>
    <au><snm>Butts</snm><fnm>CT</fnm></au>
    <au><snm>Goodreau</snm><fnm>SM</fnm></au>
    <au><snm>Morris</snm><fnm>M</fnm></au>
  </aug>
  <source>Journal of statistical software</source>
  <publisher>NIH Public Access</publisher>
  <pubdate>2008</pubdate>
  <volume>24</volume>
  <issue>3</issue>
  <fpage>nihpa54860</fpage>
</bibl>

<bibl id="B8">
  <title><p>A random graph model for massive graphs</p></title>
  <aug>
    <au><snm>Aiello</snm><fnm>W</fnm></au>
    <au><snm>Chung</snm><fnm>F</fnm></au>
    <au><snm>Lu</snm><fnm>L</fnm></au>
  </aug>
  <source>Proceedings of the thirty-second annual ACM symposium on Theory of
  computing</source>
  <pubdate>2000</pubdate>
  <fpage>171</fpage>
  <lpage>-180</lpage>
</bibl>

<bibl id="B9">
  <title><p>Stochastic blockmodels and community structure in
  networks</p></title>
  <aug>
    <au><snm>Karrer</snm><fnm>B</fnm></au>
    <au><snm>Newman</snm><fnm>ME</fnm></au>
  </aug>
  <source>Physical review E</source>
  <publisher>APS</publisher>
  <pubdate>2011</pubdate>
  <volume>83</volume>
  <issue>1</issue>
  <fpage>016107</fpage>
</bibl>

<bibl id="B10">
  <title><p>Exponential random graph models for networks with community
  structure</p></title>
  <aug>
    <au><snm>Fronczak</snm><fnm>P</fnm></au>
    <au><snm>Fronczak</snm><fnm>A</fnm></au>
    <au><snm>Bujok</snm><fnm>M</fnm></au>
  </aug>
  <source>Physical Review E</source>
  <publisher>APS</publisher>
  <pubdate>2013</pubdate>
  <volume>88</volume>
  <issue>3</issue>
  <fpage>032810</fpage>
</bibl>

<bibl id="B11">
  <title><p>Reconstructing mesoscale network structures</p></title>
  <aug>
    <au><snm>Jeude</snm><fnm>J</fnm></au>
    <au><snm>Di Clemente</snm><fnm>R</fnm></au>
    <au><snm>Caldarelli</snm><fnm>G</fnm></au>
    <au><snm>Saracco</snm><fnm>F</fnm></au>
    <au><snm>Squartini</snm><fnm>T</fnm></au>
  </aug>
  <source>Complexity</source>
  <publisher>Hindawi</publisher>
  <pubdate>2019</pubdate>
  <volume>2019</volume>
</bibl>

<bibl id="B12">
  <title><p>Community structure and scale-free collections of
  Erd{\H{o}}s-R{\'e}nyi graphs</p></title>
  <aug>
    <au><snm>Seshadhri</snm><fnm>C</fnm></au>
    <au><snm>Kolda</snm><fnm>TG</fnm></au>
    <au><snm>Pinar</snm><fnm>A</fnm></au>
  </aug>
  <source>Physical Review E</source>
  <publisher>APS</publisher>
  <pubdate>2012</pubdate>
  <volume>85</volume>
  <issue>5</issue>
  <fpage>056109</fpage>
</bibl>

<bibl id="B13">
  <title><p>A random graph model for power law graphs</p></title>
  <aug>
    <au><snm>Aiello</snm><fnm>W</fnm></au>
    <au><snm>Chung</snm><fnm>F</fnm></au>
    <au><snm>Lu</snm><fnm>L</fnm></au>
  </aug>
  <source>Experimental Mathematics</source>
  <publisher>Taylor \& Francis</publisher>
  <pubdate>2001</pubdate>
  <volume>10</volume>
  <issue>1</issue>
  <fpage>53</fpage>
  <lpage>-66</lpage>
</bibl>

<bibl id="B14">
  <title><p>R-MAT: A recursive model for graph mining</p></title>
  <aug>
    <au><snm>Chakrabarti</snm><fnm>D</fnm></au>
    <au><snm>Zhan</snm><fnm>Y</fnm></au>
    <au><snm>Faloutsos</snm><fnm>C</fnm></au>
  </aug>
  <source>Proceedings of the 2004 SIAM International Conference on Data
  Mining</source>
  <pubdate>2004</pubdate>
  <fpage>442</fpage>
  <lpage>-446</lpage>
</bibl>

<bibl id="B15">
  <title><p>Stochastic kronecker graphs</p></title>
  <aug>
    <au><snm>Mahdian</snm><fnm>M</fnm></au>
    <au><snm>Xu</snm><fnm>Y</fnm></au>
  </aug>
  <source>International Workshop on Algorithms and Models for the
  Web-Graph</source>
  <pubdate>2007</pubdate>
  <fpage>179</fpage>
  <lpage>-186</lpage>
</bibl>

<bibl id="B16">
  <title><p>Multifractal network generator</p></title>
  <aug>
    <au><snm>Palla</snm><fnm>G.</fnm></au>
    <au><snm>Lov{\'a}sz</snm><fnm>L.</fnm></au>
    <au><snm>Vicsek</snm><fnm>T.</fnm></au>
  </aug>
  <source>Proceedings of the National Academy of Sciences</source>
  <publisher>National Acad Sciences</publisher>
  <pubdate>2010</pubdate>
  <volume>107</volume>
  <issue>17</issue>
  <fpage>7640</fpage>
</bibl>

<bibl id="B17">
  <title><p>Networks: An Introduction</p></title>
  <aug>
    <au><snm>Newman</snm><fnm>M</fnm></au>
  </aug>
  <publisher>New York, NY, USA: Oxford University Press, Inc.</publisher>
  <pubdate>2010</pubdate>
</bibl>

<bibl id="B18">
  <title><p>Generating Constrained Random Graphs Using Multiple Edge
  Switches</p></title>
  <aug>
    <au><snm>Tabourier</snm><fnm>L</fnm></au>
    <au><snm>Roth</snm><fnm>C</fnm></au>
    <au><snm>Cointet</snm><fnm>JP</fnm></au>
  </aug>
  <source>J. Exp. Algorithmics</source>
  <publisher>New York, NY, USA: ACM</publisher>
  <pubdate>2011</pubdate>
  <volume>16</volume>
  <fpage>1.7:1.1</fpage>
  <lpage>-1.7:1.15</lpage>
  <url>http://doi.acm.org/10.1145/1963190.2063515</url>
</bibl>

<bibl id="B19">
  <title><p>A Markov chain Monte Carlo method for generating random (0,
  1)-matrices with given marginals</p></title>
  <aug>
    <au><snm>Rao</snm><fnm>AR</fnm></au>
    <au><snm>Jana</snm><fnm>R</fnm></au>
    <au><snm>Bandyopadhyay</snm><fnm>S</fnm></au>
  </aug>
  <source>Sankhy{\=a}: The Indian Journal of Statistics, Series A</source>
  <publisher>JSTOR</publisher>
  <pubdate>1996</pubdate>
  <fpage>225</fpage>
  <lpage>-242</lpage>
</bibl>

<bibl id="B20">
  <title><p>Relaxation-based coarsening and multiscale graph
  organization</p></title>
  <aug>
    <au><snm>Ron</snm><fnm>D</fnm></au>
    <au><snm>Safro</snm><fnm>I</fnm></au>
    <au><snm>Brandt</snm><fnm>A</fnm></au>
  </aug>
  <source>Multiscale Modeling \& Simulation</source>
  <publisher>SIAM</publisher>
  <pubdate>2011</pubdate>
  <volume>9</volume>
  <issue>1</issue>
  <fpage>407</fpage>
  <lpage>-423</lpage>
</bibl>

<bibl id="B21">
  <title><p>A multilevel bilinear programming algorithm for the vertex
  separator problem</p></title>
  <aug>
    <au><snm>Hager</snm><fnm>WW</fnm></au>
    <au><snm>Hungerford</snm><fnm>JT</fnm></au>
    <au><snm>Safro</snm><fnm>I</fnm></au>
  </aug>
  <source>Computational Optimization and Applications</source>
  <publisher>Springer</publisher>
  <pubdate>2018</pubdate>
  <volume>69</volume>
  <issue>1</issue>
  <fpage>189</fpage>
  <lpage>-223</lpage>
</bibl>

<bibl id="B22">
  <title><p>Multilevel algorithms for linear ordering problems</p></title>
  <aug>
    <au><snm>Safro</snm><fnm>I</fnm></au>
    <au><snm>Ron</snm><fnm>D</fnm></au>
    <au><snm>Brandt</snm><fnm>A</fnm></au>
  </aug>
  <source>ACM Journal of Experimental Algorithmics</source>
  <pubdate>2008</pubdate>
  <volume>13</volume>
</bibl>

<bibl id="B23">
  <title><p>Graph minimum linear arrangement by multilevel weighted edge
  contractions</p></title>
  <aug>
    <au><snm>Safro</snm><fnm>I</fnm></au>
    <au><snm>Ron</snm><fnm>D</fnm></au>
    <au><snm>Brandt</snm><fnm>A</fnm></au>
  </aug>
  <source>Journal of Algorithms</source>
  <publisher>Elsevier</publisher>
  <pubdate>2006</pubdate>
  <volume>60</volume>
  <issue>1</issue>
  <fpage>24</fpage>
  <lpage>-41</lpage>
</bibl>

<bibl id="B24">
  <title><p>Multiscale approach for the network compression-friendly
  ordering</p></title>
  <aug>
    <au><snm>Safro</snm><fnm>I</fnm></au>
    <au><snm>Temkin</snm><fnm>B</fnm></au>
  </aug>
  <source>J. Discrete Algorithms</source>
  <pubdate>2011</pubdate>
  <volume>9</volume>
  <issue>2</issue>
  <fpage>190</fpage>
  <lpage>202</lpage>
</bibl>

<bibl id="B25">
  <title><p>Networkit: An interactive tool suite for high-performance network
  analysis</p></title>
  <aug>
    <au><snm>Staudt</snm><fnm>C</fnm></au>
    <au><snm>Sazonovs</snm><fnm>A</fnm></au>
    <au><snm>Meyerhenke</snm><fnm>H</fnm></au>
  </aug>
  <source>CoRR, abs/1403.3005</source>
  <pubdate>2014</pubdate>
</bibl>

<bibl id="B26">
  <title><p>A census of planar maps</p></title>
  <aug>
    <au><snm>Tutte</snm><fnm>WT</fnm></au>
  </aug>
  <source>Canad. J. Math</source>
  <pubdate>1963</pubdate>
  <volume>15</volume>
  <issue>2</issue>
  <fpage>249</fpage>
  <lpage>-271</lpage>
</bibl>

<bibl id="B27">
  <title><p>Fast generation of planar graphs</p></title>
  <aug>
    <au><snm>Brinkmann</snm><fnm>G</fnm></au>
    <au><snm>McKay</snm><fnm>BD</fnm></au>
    <au><cnm>others</cnm></au>
  </aug>
  <source>MATCH Commun. Math. Comput. Chem</source>
  <pubdate>2007</pubdate>
  <volume>58</volume>
  <issue>2</issue>
  <fpage>323</fpage>
  <lpage>-357</lpage>
</bibl>

<bibl id="B28">
  <title><p>Program Fullgen-a program for generating nonisomorphic
  fullerenes</p></title>
  <aug>
    <au><snm>Brinkmann</snm><fnm>G</fnm></au>
  </aug>
  <source>see http://cs. anu. edu. au/bdm/plantri</source>
  <pubdate>2011</pubdate>
</bibl>

<bibl id="B29">
  <title><p>The random planar graph</p></title>
  <aug>
    <au><snm>Denise</snm><fnm>A</fnm></au>
    <au><snm>Vasconcellos</snm><fnm>M</fnm></au>
    <au><snm>Welsh</snm><fnm>DJ</fnm></au>
  </aug>
  <source>Congressus numerantium</source>
  <publisher>UTILITAS MATHEMATICA PUBLISHING INC</publisher>
  <pubdate>1996</pubdate>
  <fpage>61</fpage>
  <lpage>-80</lpage>
</bibl>

<bibl id="B30">
  <title><p>Triangle: Engineering a 2D quality mesh generator and Delaunay
  triangulator</p></title>
  <aug>
    <au><snm>Shewchuk</snm><fnm>JR</fnm></au>
  </aug>
  <source>Applied computational geometry towards geometric engineering</source>
  <publisher>Springer</publisher>
  <pubdate>1996</pubdate>
  <fpage>203</fpage>
  <lpage>-222</lpage>
</bibl>

<bibl id="B31">
  <title><p>A Delaunay refinement algorithm for quality 2-dimensional mesh
  generation</p></title>
  <aug>
    <au><snm>Ruppert</snm><fnm>J</fnm></au>
  </aug>
  <source>Journal of algorithms</source>
  <publisher>Elsevier</publisher>
  <pubdate>1995</pubdate>
  <volume>18</volume>
  <issue>3</issue>
  <fpage>548</fpage>
  <lpage>-585</lpage>
</bibl>

<bibl id="B32">
  <title><p>Random plane networks</p></title>
  <aug>
    <au><snm>Gilbert</snm><fnm>EN</fnm></au>
  </aug>
  <source>Journal of the Society for Industrial and Applied
  Mathematics</source>
  <publisher>SIAM</publisher>
  <pubdate>1961</pubdate>
  <volume>9</volume>
  <issue>4</issue>
  <fpage>533</fpage>
  <lpage>-543</lpage>
</bibl>

<bibl id="B33">
  <title><p>{On random graphs, I}</p></title>
  <aug>
    <au><snm>Erd\"{o}s</snm><fnm>P.</fnm></au>
    <au><snm>R\'{e}nyi</snm><fnm>A.</fnm></au>
  </aug>
  <source>Publicationes Mathematicae (Debrecen)</source>
  <pubdate>1959</pubdate>
  <volume>6</volume>
  <fpage>290</fpage>
  <lpage>-297</lpage>
</bibl>

<bibl id="B34">
  <title><p>On the number of edges in random planar graphs</p></title>
  <aug>
    <au><snm>Gerke</snm><fnm>S</fnm></au>
    <au><snm>McDiarmid</snm><fnm>C</fnm></au>
  </aug>
  <source>Combinatorics, Probability and Computing</source>
  <publisher>Cambridge University Press</publisher>
  <pubdate>2004</pubdate>
  <volume>13</volume>
  <issue>2</issue>
  <fpage>165</fpage>
  <lpage>-183</lpage>
</bibl>

<bibl id="B35">
  <title><p>Random planar graphs</p></title>
  <aug>
    <au><snm>McDiarmid</snm><fnm>C</fnm></au>
    <au><snm>Steger</snm><fnm>A</fnm></au>
    <au><snm>Welsh</snm><fnm>DJ</fnm></au>
  </aug>
  <source>Journal of Combinatorial Theory, Series B</source>
  <publisher>Elsevier</publisher>
  <pubdate>2005</pubdate>
  <volume>93</volume>
  <issue>2</issue>
  <fpage>187</fpage>
  <lpage>-205</lpage>
</bibl>

<bibl id="B36">
  <title><p>Streetgen: In-base procedural-based road generation</p></title>
  <aug>
    <au><snm>Cura</snm><fnm>R</fnm></au>
    <au><snm>Perret</snm><fnm>J</fnm></au>
    <au><snm>Paparoditis</snm><fnm>N</fnm></au>
  </aug>
  <source>ISPRS Annals of the Photogrammetry, Remote Sensing and Spatial
  Information Sciences</source>
  <publisher>Copernicus GmbH</publisher>
  <pubdate>2015</pubdate>
  <volume>2</volume>
  <fpage>409</fpage>
</bibl>

<bibl id="B37">
  <title><p>Generating random topology power grids</p></title>
  <aug>
    <au><snm>Wang</snm><fnm>Z</fnm></au>
    <au><snm>Thomas</snm><fnm>RJ</fnm></au>
    <au><snm>Scaglione</snm><fnm>A</fnm></au>
  </aug>
  <source>Hawaii International Conference on System Sciences, Proceedings of
  the 41st Annual</source>
  <pubdate>2008</pubdate>
  <fpage>183</fpage>
  <lpage>-183</lpage>
</bibl>

<bibl id="B38">
  <title><p>Automatic generation of water distribution systems based on GIS
  data</p></title>
  <aug>
    <au><snm>Sitzenfrei</snm><fnm>R</fnm></au>
    <au><snm>M{\"o}derl</snm><fnm>M</fnm></au>
    <au><snm>Rauch</snm><fnm>W</fnm></au>
  </aug>
  <source>Environmental modelling \& software</source>
  <publisher>Elsevier</publisher>
  <pubdate>2013</pubdate>
  <volume>47</volume>
  <fpage>138</fpage>
  <lpage>-147</lpage>
</bibl>

<bibl id="B39">
  <title><p>WaterNetGen: an EPANET extension for automatic water distribution
  network models generation and pipe sizing</p></title>
  <aug>
    <au><snm>Muranho</snm><fnm>J</fnm></au>
    <au><snm>Ferreira</snm><fnm>A</fnm></au>
    <au><snm>Sousa</snm><fnm>J</fnm></au>
    <au><snm>Gomes</snm><fnm>A</fnm></au>
    <au><snm>Marques</snm><fnm>AS</fnm></au>
  </aug>
  <source>Water science and technology: water supply</source>
  <publisher>IWA Publishing</publisher>
  <pubdate>2012</pubdate>
  <volume>12</volume>
  <issue>1</issue>
  <fpage>117</fpage>
  <lpage>-123</lpage>
</bibl>

<bibl id="B40">
  <title><p>EPANET 2: users manual</p></title>
  <aug>
    <au><snm>Rossman</snm><fnm>LA</fnm></au>
    <au><cnm>others</cnm></au>
  </aug>
  <publisher>US Environmental Protection Agency. Office of Research and
  Development. National Risk Management Research Laboratory</publisher>
  <pubdate>2000</pubdate>
</bibl>

<bibl id="B41">
  <title><p>Generating scaled replicas of real-world complex
  networks</p></title>
  <aug>
    <au><snm>Staudt</snm><fnm>CL</fnm></au>
    <au><snm>Hamann</snm><fnm>M</fnm></au>
    <au><snm>Safro</snm><fnm>I</fnm></au>
    <au><snm>Gutfraind</snm><fnm>A</fnm></au>
    <au><snm>Meyerhenke</snm><fnm>H</fnm></au>
  </aug>
  <source>International Workshop on Complex Networks and their
  Applications</source>
  <pubdate>2016</pubdate>
  <fpage>17</fpage>
  <lpage>-28</lpage>
</bibl>

<bibl id="B42">
  <title><p>Chapter 1 : Multigrid solvers and multilevel optimization
  strategies</p></title>
  <aug>
    <au><snm>Brandt</snm><fnm>A.</fnm></au>
    <au><snm>Ron</snm><fnm>D.</fnm></au>
  </aug>
  <source>Multilevel Optimization and VLSICAD</source>
  <publisher>Kluwer</publisher>
  <editor>J. Cong and J. R. Shinnerl</editor>
  <pubdate>2003</pubdate>
</bibl>

<bibl id="B43">
  <title><p>Kuratowski's theorem</p></title>
  <aug>
    <au><snm>Thomassen</snm><fnm>C</fnm></au>
  </aug>
  <source>Journal of Graph Theory</source>
  <publisher>Wiley Online Library</publisher>
  <pubdate>1981</pubdate>
  <volume>5</volume>
  <issue>3</issue>
  <fpage>225</fpage>
  <lpage>-241</lpage>
</bibl>

<bibl id="B44">
  <title><p>The battle of the water sensor networks (BWSN): A design challenge
  for engineers and algorithms</p></title>
  <aug>
    <au><snm>Ostfeld</snm><fnm>A</fnm></au>
    <au><snm>Uber</snm><fnm>JG</fnm></au>
    <au><snm>Salomons</snm><fnm>E</fnm></au>
    <au><snm>Berry</snm><fnm>JW</fnm></au>
    <au><snm>Hart</snm><fnm>WE</fnm></au>
    <au><snm>Phillips</snm><fnm>CA</fnm></au>
    <au><snm>Watson</snm><fnm>JP</fnm></au>
    <au><snm>Dorini</snm><fnm>G</fnm></au>
    <au><snm>Jonkergouw</snm><fnm>P</fnm></au>
    <au><snm>Kapelan</snm><fnm>Z</fnm></au>
    <au><cnm>others</cnm></au>
  </aug>
  <source>Journal of Water Resources Planning and Management</source>
  <publisher>American Society of Civil Engineers</publisher>
  <pubdate>2008</pubdate>
  <volume>134</volume>
  <issue>6</issue>
  <fpage>556</fpage>
  <lpage>-568</lpage>
</bibl>

<bibl id="B45">
  <title><p>Community structure in large networks: Natural cluster sizes and
  the absence of large well-defined clusters</p></title>
  <aug>
    <au><snm>Leskovec</snm><fnm>J</fnm></au>
    <au><snm>Lang</snm><fnm>KJ</fnm></au>
    <au><snm>Dasgupta</snm><fnm>A</fnm></au>
    <au><snm>Mahoney</snm><fnm>MW</fnm></au>
  </aug>
  <source>Internet Mathematics</source>
  <publisher>Taylor \& Francis</publisher>
  <pubdate>2009</pubdate>
  <volume>6</volume>
  <issue>1</issue>
  <fpage>29</fpage>
  <lpage>-123</lpage>
</bibl>

<bibl id="B46">
  <title><p>{SNAP Datasets}: {Stanford} Large Network Dataset
  Collection</p></title>
  <aug>
    <au><snm>Leskovec</snm><fnm>J</fnm></au>
    <au><snm>Krevl</snm><fnm>A</fnm></au>
  </aug>
  <source>\url{http://snap.stanford.edu/data}</source>
  <pubdate>2014</pubdate>
</bibl>

<bibl id="B47">
  <title><p>University of {F}lorida {S}parse {M}atrix {C}ollection</p></title>
  <aug>
    <au><snm>Davis</snm><fnm>T.</fnm></au>
  </aug>
  <source>NA Digest</source>
  <pubdate>1997</pubdate>
  <volume>97</volume>
  <issue>23</issue>
  <url>http://www.cise.ufl.edu/research/sparse/sparse</url>
</bibl>

<bibl id="B48">
  <title><p>The Open Graph Drawing Framework (OGDF).</p></title>
  <aug>
    <au><snm>Chimani</snm><fnm>M</fnm></au>
    <au><snm>Gutwenger</snm><fnm>C</fnm></au>
    <au><snm>J{\"u}nger</snm><fnm>M</fnm></au>
    <au><snm>Klau</snm><fnm>GW</fnm></au>
    <au><snm>Klein</snm><fnm>K</fnm></au>
    <au><snm>Mutzel</snm><fnm>P</fnm></au>
  </aug>
  <source>Handbook of Graph Drawing and Visualization</source>
  <pubdate>2013</pubdate>
  <volume>2011</volume>
  <fpage>543</fpage>
  <lpage>-569</lpage>
</bibl>

<bibl id="B49">
  <title><p>Algebraic Distance on Graphs</p></title>
  <aug>
    <au><snm>Chen</snm><fnm>J</fnm></au>
    <au><snm>Safro</snm><fnm>I</fnm></au>
  </aug>
  <source>SIAM J. Scientific Computing</source>
  <pubdate>2011</pubdate>
  <volume>33</volume>
  <issue>6</issue>
  <fpage>3468</fpage>
  <lpage>3490</lpage>
</bibl>

<bibl id="B50">
  <title><p>Single-and multi-level network sparsification by algebraic
  distance</p></title>
  <aug>
    <au><snm>John</snm><fnm>E</fnm></au>
    <au><snm>Safro</snm><fnm>I</fnm></au>
  </aug>
  <source>Journal of Complex Networks</source>
  <publisher>Oxford University Press</publisher>
  <pubdate>2016</pubdate>
  <volume>5</volume>
  <issue>3</issue>
  <fpage>352</fpage>
  <lpage>-388</lpage>
</bibl>

<bibl id="B51">
  <title><p>Advanced coarsening schemes for graph partitioning</p></title>
  <aug>
    <au><snm>Safro</snm><fnm>I</fnm></au>
    <au><snm>Sanders</snm><fnm>P</fnm></au>
    <au><snm>Schulz</snm><fnm>C</fnm></au>
  </aug>
  <source>ACM Journal of Experimental Algorithmics (JEA)</source>
  <publisher>ACM</publisher>
  <pubdate>2015</pubdate>
  <volume>19</volume>
  <fpage>2</fpage>
  <lpage>-2</lpage>
</bibl>

<bibl id="B52">
  <title><p>Recent advances in graph partitioning</p></title>
  <aug>
    <au><snm>Bulu{\c{c}}</snm><fnm>A</fnm></au>
    <au><snm>Meyerhenke</snm><fnm>H</fnm></au>
    <au><snm>Safro</snm><fnm>I</fnm></au>
    <au><snm>Sanders</snm><fnm>P</fnm></au>
    <au><snm>Schulz</snm><fnm>C</fnm></au>
  </aug>
  <source>Algorithm Engineering: Selected Results and Surveys</source>
  <publisher>Springer</publisher>
  <pubdate>2016</pubdate>
  <fpage>117</fpage>
  <lpage>-158</lpage>
</bibl>

</refgrp>
} 
\end{document}